\documentclass[epsfig,12pt]{article}
\usepackage{epsfig}
\usepackage{graphicx}
\usepackage{rotating}

\usepackage{latexsym}
\usepackage{amsmath}
\usepackage{amssymb}
\usepackage{relsize}
\usepackage{geometry}
\usepackage{color}
\usepackage{bm}

\def\beq{\begin{equation}}
\def\eeq{\end{equation}}
\def\beqn{\begin{eqnarray}}
\def\eeqn{\end{eqnarray}}

\newcommand{\ntwo}{${\mathcal N}=2\,$}
\newcommand{\none}{${\mathcal N}=1\,$}
\newcommand{\ntt}{${\mathcal N}=(2,2)\,$}
\newcommand{\nzt}{${\mathcal N}=(0,2)\,$}
\newcommand{\cpn}{CP$(N-1)\,$}
\newcommand{\ca}{{\mathcal A}}
\newcommand{\cell}{{\mathcal L}}
\newcommand{\cw}{{\mathcal W}}
\newcommand{\cs}{{\mathcal S}}

\newcommand{\pt}{\partial}

\newcommand{\zn}{$Z_N$}

\newcommand{\cde}{{\mathcal D}}
\newcommand{\cf}{${\mathcal F}$}
\newcommand{\cfe}{{\mathcal F}}

\newcommand{\gsim}{\lower.7ex\hbox{$
\;\stackrel{\textstyle>}{\sim}\;$}}
\newcommand{\lsim}{\lower.7ex\hbox{$
\;\stackrel{\textstyle<}{\sim}\;$}}

\renewcommand{\theequation}{\thesection.\arabic{equation}}

\def\beqn{\begin{eqnarray}}
\def\eeqn{\end{eqnarray}}

\def\beq{\begin{equation}}
\def\eeq{\end{equation}}
\def\ba{\beq\new\begin{array}{c}}
\def\ea{\end{array}\eeq}

\newcommand{\ntwot}{${\mathcal N}= \left(2,2\right) $ }
\newcommand{\ntwoo}{${\mathcal N}= \left(0,2\right) $ }
\renewcommand{\theequation}{\thesection.\arabic{equation}}

\newcommand{\p}{\partial}
\newcommand{\wt}{\widetilde}
\newcommand{\ov}{\overline}
\newcommand{\mc}[1]{\mathcal{#1}}

\newcommand{\lgr}{\left\lgroup}
\newcommand{\rgr}{\right\rgroup}

\def\slashed#1{\setbox0=\hbox{$#1$}             
   \dimen0=\wd0                                 
   \setbox1=\hbox{/} \dimen1=\wd1               
   \ifdim\dimen0>\dimen1                        
      \rlap{\hbox to \dimen0{\hfil/\hfil}}      
      #1                                        
   \else                                        
      \rlap{\hbox to \dimen1{\hfil$#1$\hfil}}   
      /                                         
   \fi}                                        %

\newcommand{\bxir}{\ov{\xi}{}_R}
\newcommand{\bxil}{\ov{\xi}{}_L}
\newcommand{\xir}{\xi_R}
\newcommand{\xil}{\xi_L}

\newcommand{\bzr}{\ov{\zeta}{}_R}
\newcommand{\zr}{\zeta_R}

\newcommand{\nbar}{\ov{n}}

\newcommand{\lal}{\lambda_L}

\newcommand{\blal}{\ov{\lambda}{}_L}

\newcommand{\bpsi}{\ov{\psi}{}}
\newcommand{\bphi}{\ov{\phi}{}}
\newcommand{\bxi}{\ov{\xi}{}}

\newcommand{\bi}{{\bar \imath}}
\newcommand{\bj}{{\bar \jmath}}

\newcommand{\nz}{{n^{(0)}}}
\newcommand{\no}{{n^{(1)}}}
\newcommand{\bnz}{{\ov{n}{}^{(0)}}}

\newcommand{\Dz}{{D^{(0)}}}
\newcommand{\Do}{{D^{(1)}}}

\newcommand{\sigz}{{\sigma^{(0)}}}
\newcommand{\sigo}{{\sigma^{(1)}}}
\newcommand{\bsigz}{{\ov{\sigma}{}^{(0)}}}

\newcommand{\bren}{{\beta_\text{ren}}}

\newcommand{\ssm}{{\scriptscriptstyle(M)}}
\newcommand{\sse}{{\scriptscriptstyle(E)}}

\begin{document}

\hyphenation{con-fi-ning}
\hyphenation{Cou-lomb}
\hyphenation{Yan-ki-e-lo-wicz}


\begin{titlepage}

\begin{flushright}
FTPI-MINN-09/45, UMN-TH-2828/09\\
\end{flushright}


\begin{center}
{  \Large \bf  Large-\boldmath{$N$} Solution of the Heterotic\\[2mm]
 CP\boldmath{$(N-1)$} Model with Twisted Masses}
\end{center}

\vspace{2mm}

\begin{center}

 {\large
 \bf   Pavel A.~Bolokhov$^{\,a,b}$,  Mikhail Shifman$^{\,c}$ and \bf Alexei Yung$^{\,\,c,d}$}
\end {center}

\begin{center}

$^a${\it Theoretical Physics Department, St.Petersburg State University, Ulyanovskaya~1, 
	 Peterhof, St.Petersburg, 198504, Russia}\\
$^b${\it Department of Physics and Astronomy, University of Victoria,\\
    Victoria, BC, V8P 1A1 Canada}\\
$^c${\it  William I. Fine Theoretical Physics Institute,
University of Minnesota,
Minneapolis, MN 55455, USA}\\
$^{d}${\it Petersburg Nuclear Physics Institute, Gatchina, St. Petersburg
188300, Russia
}
\end{center}

\begin{center}
{\large\bf Abstract}
\end{center}

\hspace{0.3cm}
We address a number of unanswered questions in the \nzt-deformed \cpn model with
twisted masses.  In particular, we complete the program of solving \cpn model with twisted masses in the large-$N$
limit. In hep-th/0512153 nonsupersymmetric version of the model with the $Z_N$
symmetric twisted masses was analyzed in the framework of Witten's method. In arXiv:0803.0698 
this analysis was extended:  the  large-$N$ solution of the heterotic \nzt \cpn model with no twisted masses was found. Here we solve this  model with the twisted
masses switched on. Dynamical scenarios at large and small $m$ are studied
($m$ is the twisted mass scale). We found three distinct phases and two phase transitions on the $m$ plane. 
Two phases with the spontaneously broken  \zn-symmetry are separated by a phase with unbroken \zn. This latter phase is characterized by a unique vacuum and confinement of all U(1) charged fields (``quarks").
In the broken phases (one of them is at strong coupling) there are $N$ degenerate vacua
and no confinement, similarly to the situation in the \ntt model.
Supersymmetry is spontaneously broken everywhere except a circle $|m|=\Lambda$ in the \zn-unbroken phase. 

Related issues are considered. In particular, we discuss the mirror representation for the heterotic model in a certain limiting case.

\vspace{2cm}

\end{titlepage}

\newpage

\tableofcontents

\newpage

\section{Introduction}
\setcounter{equation}{0}

Two-dimensional  CP$(N-1)$ models with twisted masses
emerged as effective low-energy theories on
the worldsheet of non-Abelian strings in a class of 
four-dimensional \ntwo\, gauge theories with unequal (s)quark 
masses~\cite{HT1,ABEKY,SYmon,HT2} (for reviews see  \cite{Trev}). 
Deforming these models in various ways (i.e. breaking supersymmetry down to \none and to nothing)
one arrives at nonsupersymmetric or heterotic \cpn models.\footnote{
Strictly speaking, the full derivation of the
heterotic \cpn model with twisted masses, valid for arbitrary values of
the deformation parameters,  from the microscopic bulk theory, is still absent.} These two-dimensional models are very 
interesting on their own, since they exhibit nontrivial dynamics with or without phase transitions 
as one varies the twisted mass scale.
In this paper we will present the large-$N$ solution of the \nzt \mbox{\cpn}  model with twisted
masses. As a warm up exercise
we first analyze this model in the limit of vanishing heterotic deformation,
i.e. the \ntt \cpn model with twisted masses (at $N\to\infty$). The majority of results presented
in this part of the paper are of a review nature
and  can be found in  \cite{HaHo,Dor,MR1,Ferrari}. In particular, Ref.~\cite{Ferrari}
deals with the large-$N$ limit in the \ntwot twisted mass deformed \cpn model.
Our solution  of the undeformed model exhibits two regimes -- the strong coupling regime and 
the Higgs regime -- with the {\em crossover} between them. We determine and briefly discuss the
Argyres--Douglas points, an issue which was  previously addressed in \cite{Ferrari,Tadpoint}.
We find it useful to collect the known results scattered in the literature, add some new nuances and, most of all,
calibrate the basic tools to be exploited below, in the introductory part.

Then we proceed to our main goal -- the large-$N$ solution of the heterotic deformation.
 Both solutions (with and without deformation)
that we present here are based on the method developed by Witten
\cite{W79,W93} (see also \cite{dadvl}) and extended in \cite{SYhet} to include the heterotic deformation.
For certain purposes we find it convenient to invoke the mirror
representation \cite{MR1,MR2}. 

An \nzt \cpn$\times C$
model on the string world sheet in the bulk theory deformed by $\mu\ca^2$
was suggested by Edalati and Tong \cite{EdTo}.
It was derived from the bulk theory in \cite{SY1} (see also \cite{BSY1,BSY2}).
Finally, the heterotic \nzt  \cpn model with twisted masses was formulated in \cite{BSY3}.
Its derivation from the microscopic bulk theory is under way \cite{underway}.

We report a number of exciting and quite unexpected results in the heterotically deformed \cpn model
with twisted masses. The model has two adjustable parameters:
one describing the strength of the heterotic deformation, and the other, $m/\Lambda$,
sets the scale of the twisted masses. 
Dynamics of the model drastically changes as we vary the value of $m$, the parameter
defining the 
 twisted masses
$$
m_k = m \exp\left(i\frac{2\pi \,k}{N}
\right),\qquad k=0,1,2,..., N-1\,.
$$
We discover three distinct phases on the $m$ plane. In the first and the third phases,
occurring at small and large values of $|m|$, the
\zn-symmetry of the model is spontaneously broken. Correspondingly,
there are $N$ degenerate vacua and no confinement. In appearance, this is akin 
to what happens in the undeformed \ntwot model. 
However, the nature of these two phases is quite different. The first one occurs
at strong  coupling (small $|m|$) while the third one at weak coupling
(large $|m|$). In essence, the latter is the Higgs phase.
Surprisingly, at intermediate values of $|m|$ we find the Coulomb/confining phase, with unbroken
\zn-symmetry and unique vacuum. It is thoroughly investigated
and the reasons for the photon to remain massless are revealed. Moreover, we find that at $|m|=\Lambda$
(in the  the Coulomb/confining phase) the vacuum of the model is supersymmetric, while
for all other values of $m$ supersymmetry is spontaneously broken.
At small and large values of the heterotic deformation parameter our solution is analytic.
At intermediate values of this parameter it is semi-analytic:
at certain stages we have to resort to numerical calculations. 

There are two phase transitions between the three distinct phases (Fig.~\ref{fig:higgsborder}). We
thoroughly analyze these phase transitions and argue that they are of the second kind.

In addition to the large-$N$ solution we address a number of related issues. For instance, in the limit
of  small
heterotic deformation parameter we build the mirror representation for the \ntwoo CP(1) model.

The organization of the paper is as follows. In a very short Section~\ref{genera} we list 
general aspects of \cpn models.  
Section~\ref{mmod} introduces, in a brief form,
our basic heterotic \cpn model with twisted masses in the gauged formulation
most suitable for the large-$N$ solution. (A discussion of the geometric formulation,
which is also helpful in consideration of some aspects, is given in Appendix E).
Section~\ref{lnscptm} presents the large-$N$ solution of the \cpn model with twisted
masses and no heterotic deformation. Now, everything is ready for the comprehensive solution of the heterotic model.

In Sect.~\ref{hecpnsm} we add a small heterotic deformation and analyze its impact on the
 large-$N$ solution. In Sect.~\ref{hetdefld} we find the
 analytic  large-$N$ solution of the heterotic \cpn model with twisted
masses in the limit of large deformations. 
Three phases and two phase transitions are identified. Intermediate values of the heterotic deformation parameter
are studied semi-analytically and numerically.
In Sect.~\ref{moccp} we focus on the second
phase, namely, the Coulomb/confining regime. 
We explain here why, unlike two other phases, the photon remains massless.
Section~\ref{relais} is devoted to related issues
and presents new results which are not necessarily based on  large $N$.
Here we construct 
the mirror representation for the heterotic  model with small values of the deformation parameter.
 Then we show, that unlike the undeformed model, the Veneziano--Yankielowicz-like and large-$N$ effective Lagrangians
 do not produce identical results -- a 
difference appears in the second order in the deformation parameter.
  Finally we present a new (albeit incomplete)
 derivation of the curve of marginal stability in the large-$N$ limit (with $Z_N$ symmetric twisted masses).
 Appendices A and B explain the Euclidean vs. Minkowski notations and formulations. In Appendix C, for convenience of the reader, we compile and compare various definitions of the heterotic deformation parameters one can find in the literature.
 The large-$N$ scaling laws for all these definitions are summarized here.
 In Appendix D
 we discuss global symmetries of the CP$(N-1)$ 
 model with the $Z_N$-symmetric twisted masses. Finally, Appendix E gives a brief review 
 of the geometric formulation of the heterotic model. Although it is only marginally used in this paper,
 it is convenient for related studies.

Section~\ref{conclu} briefly summarizes our findings. See also Erratum.


\section{Generalities}
\label{genera}
\setcounter{equation}{0}

\ntt supersymmetric \cpn sigma model was originally constructed 
\cite{orco}
in terms of \none superfields. Somewhat later it was realized \cite{Bruno}
that 
\none supersymmetry is automatically elevated up to \ntwo
provided the target manifold of the sigma model in question is K\"ahlerian (for reviews see \cite{rev1,rev2}).
The Witten index \cite{WI} of the \cpn model is $N$, implying unbroken supersymmetry
and $N$ degenerate vacua at zero energy density.
The \cpn manifold is compact; therefore, superpotential is impossible.
One can introduce mass terms, however, through the twisted masses \cite{twisted}.
The model is asymptotically free \cite{BelPo}, a dynamical scale $\Lambda$ is generated through dimensional transmutation. If the scale of the twisted masses is much larger than $\Lambda$, the theory is at weak coupling.
Otherwise it is at strong coupling. A priori, there are $N$ distinct twisted mass parameters.
However, in the absence of the heterotic deformation one of them is unobservable
(see below). In this case the model is characterized by the coupling constant $g^2$,
the vacuum angle $\theta$ and the twisted mass parameters $m_1,\,m_2,\, ..., m_N$
with the constraint
\beq
m_1 + m_2 + ... +m_N =0\,.
\label{one}
\eeq
By introducing a heterotic deformation, generally speaking, we eliminate the above constraint.
The twisted masses are arbitrary complex parameters. Of special interest 
in some instances 
is the $Z_N$ symmetric choice
\beq
m_k = m\exp\left(\frac{2\pi \,i\, k}{N}\right)\,,\qquad k = 0,1,2, ... ,N-1\,.
\label{two}
\eeq
The set (\ref{two}) will be referred to as the $Z_N$-symmetric masses. The model 
under consideration has axial U(1)$_R$ symmetry which is broken both by the chiral anomaly
and the mass terms (see Apendix \ref{app:symm} for details). 
With the mass choice (\ref{two}) the discrete
$Z_{2N}$ subgroup of this symmetry survives. We will see later that this symmetry is an
important tool in studying phase transitions in the heterotic model. In analyzing some general aspects
we will not limit ourselves to (\ref{two}). The mass parameter $m$ in Eq.~(\ref{two})
can have an arbitrary phase. One can rotate away this phase at the price of generating a 
corresponding vacuum angle $\theta$ (which effectively makes the dynamical scale parameter $\Lambda$
complex. We will follow the convention in which $\Lambda$ is kept real, while the   phase
is ascribed to $m$ (in those issues where it is important).

With the mass choice (\ref{two}) the constraint
(\ref{one}) is automatically satisfied. Mostly in this paper we will consider the case
of real and positive $m$. Sometimes however, we will relax this condition.

Where necessary, we mark the bare coupling constant by the subscript 0 and introduce
the inverse parameter $\beta$ as follows:
\beq
\beta_0 = \frac{1}{g_0^2}\,.
\eeq
At large $N$, in the 't Hooft limit,  the parameter $\beta$ scales as $N$.

There are two equivalent languages commonly used in the description of the \cpn model:
the geometric language ascending to \cite{Bruno} (see also \cite{rev2}), and the
so-called gauged formulation ascending to \cite{W79,W93}. Both have their convenient and less convenient sides.
We will discuss both formulations although construction of the $1/N$ expansion is
more convenient within the framework of the gauged formulation. At $|m|/\Lambda \to 0$
the elementary fields of the
gauged formulation (they form an  $N$-plet) are in one-to-one correspondence with the kinks in the geometric formulation.
The multiplicity of kinks -- the fact they they enter in $N$-plets --
can be readily established \cite{adam} using the mirror representation \cite{MR1}. We will discuss this in more detail later. We will review the gauged formulation 
of the model in the next section, while the geometric formulation is presented in 
Appendix \ref{app:geom}.

\section{The model}
\label{mmod}
\setcounter{equation}{0}

In this section we will briefly review the gauged formulation of
the model on which we will base the large-$N$ solution. An alternative geometric formulation
useful for general purposes 
is presented in Appendix D.

\subsection{Gauged formulation, no heterotic deformation}

We start from the gauged formulation \cite{W79,W93} of
the \mbox{\ntt} \cpn model with twisted masses
\cite{twisted} setting the heterotic deformation coupling
$\gamma = 0$.
This formulation is built on an $N$-plet of complex scalar fields $n^i$ where $i=1,2,...,N$.
We impose the constraint
\beq
\bar n_i \,n^i = 2\beta \,.
\label{m31}
\eeq
This leaves us with $2N-1$ real bosonic degrees of freedom. To eliminate one extra degree
of freedom we impose a local U(1) invariance $n^i(x)\to e^{i\alpha(x)} n^i(x)$.
To this end we introduce a gauge field $A_\mu$ which converts the partial derivative into the
covariant one,
\beq
\partial_\mu\to \nabla_\mu \equiv \partial_\mu -i\,  A_\mu\,.
\label{m32}
\eeq
The field $A_\mu$ is auxiliary; it enters in the Lagrangian without derivatives. The kinetic term of the
$n$ fields is
\beq
\cell  = \left|\nabla_\mu n^i\right|^2\,.
\label{m33}
\eeq
The superpartner to the field $n^i$ is an $N$-plet of complex two-component spinor fields $\xi^i$,
\beq
\xi^i =\left\{\begin{array}{l}
\xi^i_R\\[2mm]
\xi^i_L
\end{array}
\right.\,,
\label{m34}
\eeq
subject to the constraint
\beq
\bar{n}_i\,\xi^i =0\,,\qquad \bar\xi_i\,n^i = 0\,.
\label{npxi}
\eeq
Needless to say, the auxiliary field $A_\mu$ has a complex scalar superpartner $\sigma$ 
and a two-component complex spinor superpartner $\lambda$; both enter without derivatives.
The full \ntt-symmetric Lagrangian is\,\footnote{This is, obviously, the Euclidean version.}
\beqn
\cell &=& 
\frac{1}{e_0^2}\left(\frac{1}{4} F_{\mu\nu}^2 +\left|\pt_\mu\sigma\right|^2 + \frac{1}{2}D^2
+\bar\lambda \, i\bar{\sigma}^\mu\pt_\mu\,\lambda
\right) + i\,D\left(\bar{n}_i n^i -2\beta
\right)
\nonumber\\[3mm]
&+&
\left|\nabla_\mu n^i\right|^2+ \bar{\xi}_i\, i\bar{\sigma}^\mu\nabla_\mu\,\xi^i
+ 2\sum_i\left|\sigma-\frac{m_i}{\sqrt 2}\right|^2\, |n^i|^2
\nonumber\\[3mm]
&+&
i\sqrt{2}\,\sum_i \left( \sigma -\frac{m_i}{\sqrt 2}\right)\bar\xi_{Ri}\, \xi^i_L 
- i\sqrt{2}\,\bar{n}_i \left(\lambda_R\xi^i_L - \lambda_L\xi^i_R \right)
\nonumber\\
&+&
i\sqrt{2}\,\sum_i \left( \bar\sigma -\frac{\bar{m}_i}{\sqrt 2}\right)\bar\xi_{Li}\, \xi^i_R 
- i\sqrt{2}\,{n}^i \left(\bar\lambda_L\bar\xi_{Ri} - \bar\lambda_R\bar\xi_{Li} \right),
\label{bee31}
\eeqn
where $m_i$ are twisted mass parameters, and the limit $e_0^2\to\infty$ is implied. 
Moreover,
\beq
\bar\sigma^\mu = \{1,\,i\sigma_3\}\,,
\label{wtpi3}
\eeq
see Appendix~\ref{app:eucl}.

It is clearly seen that the auxiliary field $\sigma$
enters in (\ref{bee31}) only through the  combination
\beq
\sigma -\frac{m_i}{\sqrt 2}\,.
\label{combi}
\eeq
By an appropriate shift of $\sigma$
one can always redefine the twisted mass parameters in such a way that the constraint
(\ref{one}) is satisfied.
The U(1) gauge symmetry is built in. This symmetry eliminates one bosonic degree of freedom, leaving us with $2N-2$ dynamical bosonic degrees of freedom inherent to CP$(N-1)$ model.

\subsection{Switching on the heterotic deformation}
\label{gfsothd}

The general formulation of \ntwoo gauge theories in two dimensions was addressed by Witten in 
\cite{W93}, see also \cite{Witten:2005px}.
In order to deform the \cpn model breaking \ntt down to \mbox{\nzt}
we must introduce a right-handed spinor field $\zeta_R$ (with a bosonic superpartner \cf)
whose  target space is $C$, which is coupled to other fields as follows
\cite{EdTo,SY1}:
\beqn
\Delta\cell 
&=&
\bar\zeta_R\, i\pt_L\, \zeta_R   +\bar\cfe\,\cfe
	\nonumber\\[2mm]
&+&
 2i\, \omega \, \bar\lambda_L\, \zeta_R +  2i\, \bar{\omega} \, \bar\zeta_R\, \lambda_L -
	 2i\, \omega \, \cfe\, \sigma -
 2i\, \bar{\omega} \, \bar\cfe\, \bar{\sigma}\,,
 \label{deforte}
\eeqn
where $\omega$ is a deformation parameter\footnote{The reader is referred to
Appendix~\ref{app:het} where definitions of heterotic deformation parameters useful in different regimes
are brought together. In this paper, only parameter $ \omega $ will be essentially exploited.}.

This term must be added to the \ntt Lagrangian (\ref{bee31}). 
It is quite obvious that the dependence on (\ref{combi}) is gone. 
The deformation term (\ref{deforte}) has a separate dependence on $\sigma$, not reducible to the
combination (\ref{combi}). Therefore, for a generic choice, all $N$ twisted mass
parameters $m_1,\,m_2,\, ..., m_N$ become observable, Eq. (\ref{one}) is no longer valid.

\vspace{1mm}

Eliminating \cf, $\bar\cfe$  and $\bar\lambda ,\,\lambda$ we get
\beq
\Delta\cell = 4\, |\omega |^2\,|\sigma |^2\,,
\label{deffpp}
\eeq
while the constraints (\ref{npxi}) are replaced by
\begin{align}
\bar{n}_i\,\xi^i_L & ~=~ 0\,,
& \bar\xi_{Li}\,n^i & ~=~ 0\,,
\nonumber\\[2mm]
\bar{n}_i\,\xi^i_R & ~=~  \sqrt{2}\,\bar\omega \bar\zeta_R\,,
& \bar\xi_{Ri}\,n^i  & ~=~  \sqrt{2}\,\omega \zeta_R\,.
\label{npxip}
\end{align}

We still have to discuss  $N$ dependence of the deformation parameter $\omega$.
We want to single out appropriate powers of $N$ so that the large-$N$ limit will be smooth.
From (\ref{deffpp}) it is clear that $\omega$ scales as
$\sqrt N$.

One can restore the original form of 
the constraints (\ref{npxi})  by   shifting  the $ \xir $ fields, namely,
\beq
\xi_R' ~~=~~ \xir ~-~ \frac{1}{2\beta}\,\sqrt{2}\,\ov{\omega}\, n\, \bar\zeta_R\,, \qquad
\bar\xi_R' ~~=~~  \bar\xi_R ~-~ \frac{1}{2\beta}\,\sqrt{2}\,\omega\, \bar{n}\, \zeta_R\,.
\label{wtpi7}
\eeq
This obviously changes the normalization of the kinetic term for $ \zr $, which we can
bring back to its canonic form by a rescaling $ \zr $,
\beq
	\zr  \to   ( 1 - |\tilde\gamma|^2 )\, \zr\,,
	\label{wtpi8}
\eeq
where the relation between $\tilde\gamma$ and $\omega$ as well as their $N$-dependence
are given in Appendix~\ref{app:het}.
As a result of these transformations,  the following Lagrangian emerges \cite{BSY3}:
\beqn
\cell 
&=&
 \bzr\, i\pt_L\, \zr  ~+~ 
  | \pt_k n |^2 ~+~
 \bxir\, i\p_L\, \xir ~+~  \bxil\, i\p_R\, \xil 
	\nonumber\\[3mm]
&+&
  \sum_l |m^l|^2 \left|n^l \right|^2 
~-~ i m^l\, \bar{\xi}_{Rl} \xi_L^l ~-~ i\bar{m}^l\, \bar{\xi}_{Ll} \xi_R^l
\nonumber\\
&+& 
\frac{1}{\sqrt{2\beta}}\left\{\rule{0mm}{5mm}
\wt{\gamma}\, i\pt_L\nbar \xir\, \zr \,+\, \ov{\wt\gamma}\, \bxir i\p_L n\, \bzr
\,+\, i\wt{\gamma}\, m^l\, \bar{n}_l \, \xi_L^l\, \zr 
	\,-\, i\ov{\wt\gamma}\, \bar{m}^l\, \bar{\xi}_{Ll} n^l\, \bzr
		\right\}
\nonumber\\[3mm]
&+&
\frac{1}{2\beta} \left\{\rule{0mm}{6mm}
  (\nbar  \p_k n)^2
	\,-\,  (\nbar  i\p_R n)  \bxil \xil \,-\,  (\nbar  i\p_L n) \bxir \xir
	\right.
\nonumber\\[3mm]
&+& (1-|\tilde\gamma|^2)\,\bxil\xir\,\bxir\xil \,-\, \bxir\xir\,\bxil\xil
\,+\, |\tilde\gamma|^2\, \bxil \xil\, \bzr \zr
	\nonumber\\[5mm]
&-&
\left.
 (1-|\tilde\gamma|^2)
	\left(
	 \left|\sum m^l |n^l|^2 \right|^2 
		- i m^l\, |n^l|^2 (\bxir\xil) - i\bar{m}^l\, |n^l|^2(\bxil\xir)
	\right) \right\}.
	\label{sigma_phys}
\eeqn
The sums over $l$ above run from $l=1$ to $N$. If the masses are chosen \zn-symmetrically, see
(\ref{two}), this Lagrangian is explicitly \zn-symmetric, see  Appendix~\ref{app:symm}.

If all $m_l$ are zero,  the model (\ref{sigma_phys}) reduces to the \nzt \cpn model
derived in \cite{SY1}, see (\ref{bee31}). 
Later on we will examine
other special choices for the  the mass terms. Here we will only note that
with all $m_l \neq 0$ the masses of the boson and fermion excitations following from
(\ref{sigma_phys}) split \cite{BSY3}. Say, in the $l_0$-th vacuum
\beqn
M_{\rm ferm}^{(l)} 
&=&
 m^l - m^{l_0} + |\tilde\gamma|^2\,m^{l_0} \,,
	\nonumber\\[2mm]
	\left| M_{\rm bos}^{(l)} \right|
	&=&
	\sqrt{ \left| M^{(l)}_{\rm ferm} \right|^2 - |\tilde\gamma|^4\, \left| m^{l_0}\right|^2}
	\,,
	\nonumber\\[3mm]
	l
	&=&
	 1,2, ..., N;\quad l\neq l_0\,.
\label{ferbosmasssplit}
\eeqn
The splitting between the boson and fermion masses shows that \ntwoo supersymmetry is
spontaneously broken, see \cite{BSY3} for further details.

	The model (\ref{sigma_phys}) still contains redundant fields.
	In particular, there are $N$ bosonic fields $n^l$ and $N$ fermionic $\xi^l$,
	whereas the number of physical degrees of 
	freedom is $2\times(N-1)$. 
	One can readily eliminate the redundant fields, say, $n^N$ and $\xi^N$,
	by exploiting the constraints (\ref{npxi}). Then explicit \zn-symmetry will be lost, of course.
	It will survive as an implicit symmetry.

\section{Large-\boldmath{$N$} solution
of the \cpn model with twisted masses}
\label{lnscptm}
\setcounter{equation}{0}

In this section we present the large-$N$ solution of the \ntwot supersymmetric \cpn model 
with twisted masses (\ref{bee31}). We consider a special case of mass deformation (\ref{two}) 
preserving the $Z_N$ symmetry of the model. 
The \ntwot  model with the vanishing twisted masses, as well as nonsupersymmetric \cpn model,
were solved by Witten in the large-$N$ limit \cite{W79}.
The same method was used in
\cite{GSYphtr} to study nonsupersymmetric \cpn 
model with twisted mass. In this section we will generalize this analysis 
to solve the \ntwot theory with twisted masses included. 

As was mentioned in the Introduction,
many issues discussed in this section were previously addressed  in 
\cite{HaHo,Dor,MR1,Ferrari}. 
The large-$N$ limit in the \ntwot supersymmetric \cpn model 
with twisted masses was treated in \cite{Ferrari}.
Moreover, the large-$N$ expansion, in fact,
 is not the {\em  only} method in the studies 
of the \ntwot supersymmetric \cpn
 model. Indeed, in this  model exact superpotentials  of the Veneziano--Yankielowicz type are known
for arbitrary $N$ \cite{AdDVecSal,ChVa,W93,HaHo,Dor}. We use the large-$N$ expansion in this
section 
to prepare tools we will exploit later to solve the \ntwoo supersymmetric \cpn model (for which no
exact superpotentials are known).

First let us very briefly review the physics of nonsupersymmetric \cpn model revealed
by the large-$N$ solution \cite{GSYphtr}.
In the limit of vanishing masses, the \mbox{\cpn} model is known to 
be a strongly coupled asymptotically free
field theory
\cite{BelPo}. A dynamical scale $\Lambda$ is generated as a result of
dimensional transmutation.
At large $N$ it can be solved by virtue of the $1/N$ expansion
\cite{W79}.
The solution  exhibits a ``composite massless photon"
coupled to $N$ quanta $n^i$, each with charge $1/\sqrt N$ with respect
to this photon. In two dimensions the corresponding Coulomb potential is long-range.
It causes linear confinement, so that only the $\bar n\,n$ pairs show up in the
spectrum \cite{Coleman,W79}. This is the reason why we will refer to this phase as
``Coulomb/confining." In the Coulomb/confining phase the vacuum 
is unique and the $Z_N$ symmetry is unbroken.

On the other hand, if the mass deformation parameter $m$ is $\gg\Lambda$,
the model is at weak coupling, the field $n$ develops a vacuum
expectation value (VEV),
there are $N$ physically equivalent vacua, in each of which the
$Z_N$ symmetry is spontaneously broken. We  refer to this regime
as the
Higgs phase.

In Ref. \cite{GSY05}
it was argued that (nonsupersymmetric) twisted mass deformed \mbox{\cpn} model
undergoes a phase transition when the value of the mass parameter is
$\sim \Lambda$, to the Higgs phase with the broken $Z_N$ symmetry.
 In \cite{GSYphtr} this result was confirmed by the explicit
 large-$N$ solution. (Previously
the issue of two phases and phase transitions in related models
was  addressed by Ferrari
\cite{Ferrari,Ferrari2}.)

In the \ntwot supersymmetric \cpn model, generally speaking,  we do not expect a phase transition
in the 
twisted mass to occur.  In this section we confirm this expectation demonstrating
that the $Z_N$ symmetry is  broken at all values of the twisted mass. 
(See, however, the end of Sect.~\ref{tscreg}.)
Still, the theory has two 
distinct regimes, the Higgs regime at large $m$ and the  strong-coupling one 
at small $m$.\footnote{At finite $N$ there is {\em no} phase transition between these regimes. Instead, one has a crossover. This is explained after Eq.~(\ref{22sigmaapp}).}

Since the action (\ref{bee31}) is quadratic in the fields $n^{i}$ and $\xi^i$
we can integrate over these fields and then minimize the resulting
effective action with respect to the  fields from the gauge multiplet. The large-$N$ limit ensures the corrections to the saddle point approximation to be  small. In fact,
this procedure boils down to calculating a small set of one-loop graphs with the
$n^{i}$ and $\xi^i$  fields propagating in loops.

In the Higgs regime the field $n^{i_0}$ develops a VEV.
One can always choose $i_0=0$ and denote $n^{i_0}\equiv n$. 
The field $n$, along with $\sigma$, are
our order parameters that distinguish between the
strong coupling  and Higgs regimes. These parameters show a rather dramatic crossover behavior
when we move from one regime to another.

Therefore, we do not want
to integrate over $n$ {\em a priori}. Instead,
we will stick to the following strategy:  we integrate over $N-1$
fields $n^{i}$ with $i \ne 0$.
The resulting effective action is to be considered as
a functional of $n^0\equiv n$, $D$ and
$\sigma$. To find the vacuum configuration, we will then minimize the
effective action with
respect to $n$, $D$ and
$\sigma$.

The fields $ n^i $ and $ \xi^i $  ($ i = 1,...\, N-1 $) enter the Lagrangian quadratically,
\beqn
\Delta\cell &=&
 \nbar{}_i 
		   \left( - \p_k^2 ~+~ \Bigl| \sqrt{2}\sigma - m^i \Bigr|^2 +  i\, D \right) n^i
		   + ...
\nonumber\\[4mm]
&+&
\left( \bxi_{Ri}\,\,  \bxi_{Li} \right)
		\left( \begin{matrix}
			i\,\pt_L  &  
			i \Bigl(\sqrt{2}\sigma -  m^i\Bigr) \\
			i \Bigl(\sqrt{2}\ov{\sigma} - \ov{m}{}^i \Bigr) &  
			i\,\p_R 
		     \end{matrix} \right)
		\left( \begin{matrix}
			\xi_R^i \\[5mm] \xi_L^i
		     \end{matrix} \right)+ ... ,
\eeqn
where the ellipses denote terms which contain neither $n$ nor $\xi$ fields.
Hence, integration over $n^{i}$ and $\xi^i$ in (\ref{bee31})
yields the following ratio of the determinants:
\beq
 \frac{
\prod_{i=1}^{N-1} {\rm det}\, \left(-\pt_{k}^2 
   + \bigl| \sqrt{2}\sigma - m_i \bigr|^2\right)}{
\prod_{i=1}^{N-1}{\rm det}\, \left(-\pt_{k}^2 +iD
   + \bigl| \sqrt{2}\sigma - m_i \bigr|^2\right)},
\rule{0mm}{10mm}
\label{det}
\eeq
where we dropped the gauge field $A_k\rule{0mm}{5mm}$ which is
irrelevant for the following determination of  vacuum structure.\footnote{Needless to say, this field is important in the spectrum calculation.} The  determinant in the denominator
comes from the
boson loops while that in the numerator from the fermion loops. Note, that the $n^{i}$ mass 
squared
 is given by $iD+|\sqrt{2}\sigma-m_i|^2$ while that of fermions $\xi^i$
is $|\sqrt{2}\sigma-m_i|^2$. If supersymmetry is unbroken (i.e.  $D=0$) these masses are equal,
and the ratio of the determinants reduces to unity, as
it should be, of course.

Calculation of the determinants in Eq.~(\ref{det}) 
is straightforward. 
We easily get the following contribution to the
effective Lagrangian:
\beqn
\Delta{\mathcal  L} 
&=&
\sum_{i=1}^{N-1}\frac{1}{4\pi}\left\{\left(iD+\bigl|\sqrt{2}\sigma-m_i\bigr|^2\right)
\lgr \ln\, {\frac{M_{\rm uv}^2}{iD + \bigl|\sqrt{2}\sigma-m_i\bigr|^2}} + 1 \rgr
\right.
\nonumber\\[4mm]
&-&
\left. 
\bigl|\sqrt{2}\sigma-m_i\bigr|^2
\lgr \ln\, {\frac{M_{\rm uv}^2}{\bigl|\sqrt{2}\sigma-m_i\bigr|^2}} + 1\rgr
\right\},
\label{detr}
\eeqn
where quadratically divergent contributions from bosons and fermions do
not depend on
$D$ and $\sigma$ and cancel each other. Here $M_{\rm uv}$ is an ultraviolet (UV) cutoff.
 The bare coupling constant
 $2\beta_0$ in (\ref{bee31}) can be parametrized as
\beq
2\beta_0~~=~~\frac{N}{4\pi}\, \ln\, {\frac{M_{\rm uv}^2}{\Lambda^2}}\,.
\eeq
Substituting this expression in (\ref{bee31}) and adding 
the one-loop correction
(\ref{detr})
we see that the term proportional to 
$iD \,\ln {M_{\rm uv}^2}$ is canceled, and the effective action is
expressed in terms of the renormalized coupling constant,
\beq
\rule{0mm}{8mm}
	2\bren~~=~~\frac{1}{4\pi}\, 
	\sum_{i=1}^{N-1}\ln\, {\frac{iD +\bigl|\sqrt{2}\sigma-m_i\bigr|^2}{\Lambda^2}}\, .
\label{coupling}
\eeq

\vspace{2mm}

Assembling  all contributions together and dropping 
the gaugino fields $\lambda$ we get the effective potential as 
a function of $n$, $D$ and $\sigma$ fields  in
the form
\beqn
	V_\text{eff} & =& \int d^2x 
		\biggl\{  \lgr iD ~+~ \bigl|\sqrt{2}\sigma - m_0\bigr|^2 \rgr |n|^2 
	\nonumber\\[4mm]
	&-& 
	\frac{1}{4\pi}\, \sum_{i=1}^{N-1} \lgr iD ~+~ \bigl|\sqrt{2}\sigma - {m_i}\bigr|^2 \rgr\,
		\ln\, \frac{ iD \,+\, \left| \sqrt{2}\sigma - m_i \right|^2} {\Lambda^2}
\nonumber\\[4mm]
	&+& 
	\frac{1}{4\pi}\, \sum_{i=1}^{N-1} \bigl| \sqrt{2}\sigma - m_i\bigr|^2\,
			\ln\, \frac{ \bigl| \sqrt{2}\sigma -  m_i \bigr|^2 }{ \Lambda^2 }
	+
	\frac{1}{4\pi}\, iD\, (N-1)  \biggr \}  .
	\nonumber\\
	\label{Veff22}
\eeqn
Now, to find the vacua, we must minimize the 
effective potential (\ref{Veff22}) with respect to $n$, $D$ and $ \sigma $. In this way we  arrive at
the set of	the vacuum equations,
\beqn
\label{22eff1}
	&&
	|n|^2  ~~=~~ 2\,\bren\,, 
\\[3mm]				
	&&
	\lgr i D + \bigl|\sqrt{2}\sigma - m_0 |^2 \rgr n ~~=~~ 0\,, 
	\label{22eff2}
	\\[3mm]
	&&
	\bigl(\sqrt{2}\sigma - m_0 \bigr)|n|^2 
		-
	\frac{1}{4\pi} \sum_{i=1}^{N-1}\,
			\bigl(\sqrt{2}\sigma - m^i\bigr)\,
		{\rm ln}\, 
		\frac{i D + \bigl| \sqrt{2}\sigma \,-\, m^i \bigr|^2}
		{\bigl| \sqrt{2}\sigma - m^i \bigr|^2}
		 ~~=~~ 0\,,
\label{22eff3}
\nonumber\\
\eeqn
where $2\bren$ is determined by Eq.~(\ref{coupling}).

From Eq.~(\ref{22eff2}) it is obvious that there are two options: either
\beq
\label{higgsph22}
	 iD ~+~ \bigl| \sqrt{2}\sigma - m_0 \bigr|^2 ~~=~~ 0  
\eeq
or
\beq
\label{strongph22}
	 n ~~=~~ 0 \,. 
\eeq
	These two distinct solutions correspond to the Higgs and the strong-coupling regimes of the theory, respectively. 
	Equations (\ref{22eff1})--(\ref{22eff3}) represent our {\em master set} which
determines the vacua of the theory. 

\subsection{The Higgs regime}
\label{hireg}

Consider first the Higgs regime.
For large $m$ we have the solution
\beq
D~=~0,\qquad \sqrt{2}\sigma~=~m_0,\qquad |n|^2~=~2\bren \,.
\label{higgsvac}
\eeq
The first condition here, $D=0$, means that \ntwot supersymmetry is not broken and the vacuum
energy is zero. Integrating over $n$'s and $\xi$'s we fixed $n^{0}\equiv n$. Clearly,
alternatively  we could have fixed
any other $n^{i_0}$. Then, instead of (\ref{higgsvac}), we would get
\beq
D~=~0,\qquad \sqrt{2}\sigma~=~m_{i_0},\qquad |n^{i_0}|^2~=~2\bren\,,
\label{higgsvacN}
\eeq
demonstrating the  presence of $N$ degenerate vacua. The discrete chiral $Z_{2N}$ symmetry (\ref{bee35})
is broken by these VEV's down to $Z_2$. Substituting 
the above  expressions   for $D$ and $\sigma$ in (\ref{coupling})
we get the renormalized coupling 
\beq
2\bren~=~\frac{1}{4\pi}\, 
\sum_{i=1}^{N-1}\ln\, {\frac{|m_{0}-m_i|^2}{\Lambda^2}}~=~\frac{N}{2\pi}\,\ln\,{\frac{m}{\Lambda}}\, ,
\label{22higgscoupling}
\eeq
where we calculated the sum over $i$ in the large-$N$ limit for the special choice of masses
(\ref{two}).

In each vacuum there are $2(N-1)$ elementary excitations\,\footnote{Here we count
real degrees of
freedom. The action (\ref{bee31}) contains $N$ complex fields
$n^i$.
The phase of $n^{i_0}$ can be eliminated from the very beginning.
The condition $|n^i|^2 = 2\beta$ eliminates one extra field.} 
with the physical masses
\beq
M_i ~=~ |m_i-m_{i_0}|\,,\qquad i\neq i_0\,.
\label{elmass}
\eeq
In addition to the elementary excitations, there are kinks (domain ``walls" which are particles in two
dimensions) interpolating between these vacua. 
Their masses scale as
\beq
M^{\rm kink}_{i} ~\sim~ \bren\,M_i \,.
\label{kinkmass}
\eeq
The kinks  are much  heavier than elementary
excitations at weak coupling. Note that they have nothing to do
with Witten's $n$ solitons \cite{W79} identified as the $n^i$ fields at
strong coupling, see Sect.~\ref{wtnfcb}.

Since $|n^{i_0}|^2=2\bren$ is positively defined we see that the crossover point is
at $m=\Lambda$. Below this point, the VEV of the $n$ field vanishes,
 and we are in the strong coupling regime.

\subsection{The strong coupling regime}
\label{tscreg}

For small $m$ the
solutions of Eqs. (\ref{22eff1}) -- (\ref{22eff3}) can be readily found,
\beq
D~=~0,\qquad n~=0~,\qquad
2\bren~=~\frac{1}{4\pi}\, 
\sum_{i=1}^{N-1}\ln\, {\frac{\bigl|\sqrt{2}\sigma-m_i\bigr|^2}{\Lambda^2}}=0 \,.
\label{strongvac}
\eeq
Much in the same way as in the Higgs regime, the condition $D=0$ means that   \ntwot
 supersymmetry remains unbroken.
 
 Note that at large $N$, the summation in (\ref{strongvac}) can be extended to include the
 $i=0$ term,
 \beq
 2\bren~=~\frac{1}{4\pi}\, 
\sum_{i=0}^{N-1}\ln\, {\frac{\bigl|\sqrt{2}\sigma-m_i\bigr|^2}{\Lambda^2}}=0 \,,
\label{strongvacp}
 \eeq
because (as we will show below) $\sqrt{2}\sigma\sim \Lambda$ in this regime and is not
close to any of $m_i$ at $|m|<\Lambda$.

The last equation can be identically rewritten as 
\beq
\prod_{i=0}^{N-1}\left|\sqrt{2}\sigma-m_i\right| \,=\,\Lambda^N \,.
\label{Witcond}
\eeq
For the \zn-symmetric masses
Eq. (\ref{Witcond}) can be solved. Say, for even $N$ one can rewrite this equation in the form
\beq
\left|\left(\sqrt{2}\sigma\right)^N ~-~ m^N\right| ~=~ \Lambda^N \,.
\label{tumvn}
\eeq
due to the fact that with the masses given in (\ref{two})
\beqn
\sum m_i &=& 0\,,
\nonumber\\[2mm]
\sum_{i,j;\,i\neq j}\,\, m_i m_j &=& 0\,,
\nonumber\\[2mm]
&...&
\nonumber\\[2mm]
\sum_{i_1,i_2,...,i_{N-1}} m_{i_1} m_{i_2} ... m_{i_{N-1}} &=& 0\,,\qquad \left(i_1\neq i_2\neq ...\neq i_{N-1}\right).
\eeqn
Equation (\ref{tumvn})  has $N$ solutions
\beq
\sqrt{2}\sigma ~=~ \left(\Lambda^N+m^N\right)^{1/N}\,
\exp\left( \frac{2\pi\,i\, k}{N}
\right), \quad k=0, ..., N-1,
\label{22sigma}
\eeq
where we assumed for simplicity that $m\equiv m_0$ is real and positive.
(This is by no means necessary; we will relax this assumption at the end of this section.)
Note that the phase factor of $\sigma$ in (\ref{22sigma}) does not follow from (\ref{Witcond}). Rather, its emergence 
is explained by explicit breaking of the axial U(1)$_{R}$ symmetry down to
$Z_{2N}$ through the anomaly and non-zero masses (\ref{two}), see Appendix \ref{app:symm},
with the subsequent spontaneous breaking of $Z_{2N}$ down to $Z_2$. Once we have one solution to
(\ref{Witcond}) with the nonvanishing $\sigma$ we can generate all $N$ solutions (\ref{22sigma})
by the $Z_{2N}$ transformation \cite{W79}.

\begin{figure}
\centerline{\input{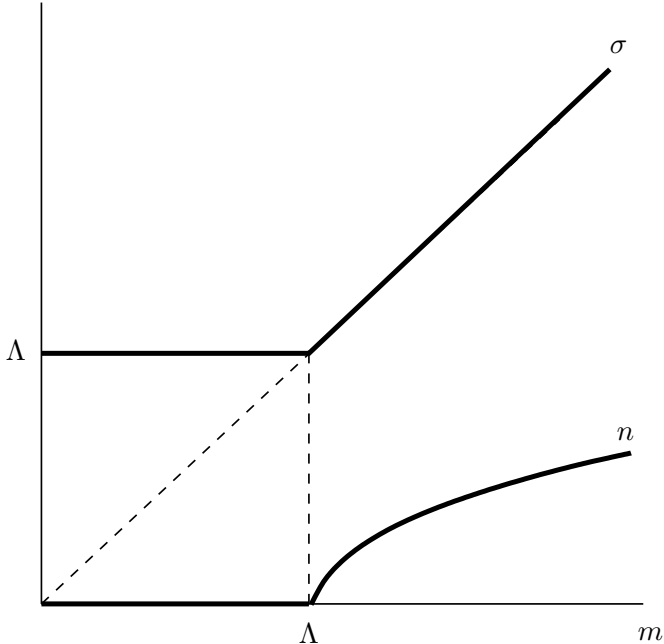}}
\caption{\small Plots of $n$ and $\sigma$ VEVs (thick lines) vs. $m$
in the \ntt \cpn model with twisted masses as in (\ref{two}). }
\label{fig22nsigma}
\end{figure}

Although we derived  Eq.~(\ref{Witcond}) in the large-$N$ approximation, the 
complexified version of this equation,
\beq
\prod_{i=0}^{N-1}\left(\sqrt{2}\sigma-m_i\right) \,=\,\Lambda^N \,,
\label{Witcondc}
\eeq
is in fact,
exact, since  this equation as well as the solution (\ref{22sigma}) follow from the
Veneziano--Yankielowicz-type  effective Lagrangian exactly derived in the \ntwot \cpn model  
in  \cite{AdDVecSal,ChVa,W93,HaHo,Dor}. The Veneziano--Yankielowicz  Lagrangian implies (\ref{Witcondc}) 
even at finite $N$.

The solution (\ref{22sigma}) shows the presence of $N$ degenerate vacua. Since $\sigma \neq 0$  in all these vacua
the discrete chiral $Z_{2N}$ symmetry is broken down to $Z_2$ in the strong-coupling regime, much in the same way as
in the Higgs regime. This should be contrasted with the large-$N$ solution of the nonsupersymmetric
massive \cpn model \cite{GSYphtr}. In the latter case, $\sigma=0$ in the strong coupling phase, 
therefore, the theory
has   a single vacuum state in which the $Z_{2N}$ symmetry is restored. This is a signal of a phase transition
separating the Higgs and Coulomb/confining phases in the nonsupersymmetric massive \cpn model \cite{GSYphtr}.

In fact, in the large-$N$ approximation the formula ({\ref{22sigma}) can be rewritten as
\beq
\sqrt{2}\sigma ~=~ \exp\left( \frac{2\pi\,i\, k}{N}
\right)\times \left\{
\begin{array}{cc}
\Lambda,\;\;\;\;\;\;m <\Lambda\\[1mm]
\;m,\;\;\;\;\;\; m > \Lambda
\end{array},
\right.
 \qquad k=0, ..., N-1
\label{22sigmaapp}
\eeq
with the exponential accuracy   $O\left(e^{-N}\right)$. Note that at large $m$ this formula reproduces our
result (\ref{higgsvacN}) obtained in the Higgs regime. In the limit $m\to 0$ it gives Witten's
result  \cite{W79}.

The  VEVs $n$ and $\sigma$ as functions of $m$ are plotted in Fig.~\ref{fig22nsigma}.
These plots suggest that we have  discontinuities in derivatives over $m$ for both order
parameters. Taken at its face value, this would signal a phase transition, of course. We note, however, that 
the exact formula
(\ref{22sigma}) shows a smooth behavior in $\sigma$. Therefore, we interpret the discontinuity in
(\ref{22sigmaapp}) as an artifact of the large-$N$ approximation. The crossover transition between 
the two regimes becomes exceedingly more pronounced as we increase 
$N$ and turns into the second-order phase transition in the
limit $N\to\infty$. We stress again that the $Z_{2N}$ symmetry is broken down to $Z_2$ in 
the both regimes.

There is one interesting special point in Eq.~(\ref{Witcondc}).
Relaxing the requirement of reality of the parameter $m_0$ we can choose the 
product\,
\beq
\prod_{i=0}^{N-1}\,(-m_i) \,=\,\Lambda^N\,.
\label{ADpoint}
\eeq
At this particular point 
 Eq.~(\ref{Witcondc}) reduces to $\sigma^N=0$ with the solution
\beq
\sigma ~=~ 0\,.
\label{ADsigma}
\eeq
All $N$ vacua coalesce!  
This is a two-dimensional ``reflection" of the four-dimensional
Argyres--Douglas (AD) point \cite{AD,APSW}.

\subsection{Generic twisted masses and  the Argyres--Douglas points}
\label{genadp}

In this subsection we briefly describe the AD points in the undeformed \ntwot \cpn model.
At these points one or more kinks interpolating between different vacua of the model become
{\em massless}. These points determine a nontrivial conformal regime in the
theory. We use the complexified version of the vacuum equation (\ref{Witcondc})
appropriate for a generic choice of the twisted mass parameters.

Let us have a closer look at this equation given a set of arbitrary masses.
Our task is to find  the values of mass parameters such that
two roots of this equation coalesce, $\sigma_1=\sigma_2$.  Near the common value of
$\sigma$ Eq.~(\ref{Witcondc}) can be simplified, namely,
\beq
\left(\sqrt{2}\sigma-m_{12}\right)^2-\frac{\Delta m^2_{12}}{4} ~=~ \Lambda_{\rm eff}^2 ~\equiv~ 
\frac{\Lambda^N}{\prod_{i\neq 1,2}(m_{12}-m_i)},
\label{sigmaeqAD}
\eeq
where
\beq
m_{12}~=~\frac12(m_1+m_2),\qquad \Delta m_{12}~=~m_1-m_2
\label{m12}
\eeq
Equation (\ref{sigmaeqAD}) gives
\beq
\sqrt{2}\,\sigma_{1,2} \,~=~\, m_{12} ~\pm~ \sqrt{\frac{\Delta m^2_{12}}{4} \,+\, \Lambda_{\rm eff}^2}.
\eeq
Two vacua coalesce if the square root vanishes,
\beq
-\Delta m^2_{12}\prod_{i\neq 1,2}(m_{12}-m_i) ~=~ 4\Lambda^N\,.
\label{AD}
\eeq
At this AD point one of $N$ kinks interpolating between the vacua at $\sigma_1$ and $\sigma_2$
becomes massless.

Similarly, one can consider more complicated AD points in which more than two vacua coalesce.
At these AD points more than one kink becomes massless. The point (\ref{ADpoint})
corresponds to a very special  regime in which  all $N$ vacua coalesce (for the \zn-symmetric choice of 
masses   on the circle (\ref{two})). At this point in the mass parameter space one of $N$ kinks interpolating between each two ``neighboring" vacua becomes massless. This AD point was
studied previously  in \cite{Tadpoint}. We remind that the \ntwot supersymmetric \cpn model
is an effective theory on the world sheet of the non-Abelian string in \ntwo SQCD
(with the U$(N)$ gauge group and $N_f=N$   flavors \cite{HT1,ABEKY,SYmon,HT2}).
Therefore, the massless kinks at the AD points in two dimensions correspond to massless confined monopoles
at the AD points in four-dimensional bulk theory.

We pause here to make a remark unrelated to the Argyres--Douglas points. Assume one has a (nearly) generic
set of twisted masses subject to a single constraint
\beq
\prod_{i=0}^{N-1}\,(-m_i)  = \Lambda^N\,.
\label{weass}
\eeq
Then Eq.~(\ref{Witcondc}) has a solution $\sigma =0$, with other $N-1$ solutions
$\sigma\neq 0$. Now, if we introduce the heterotic deformation $\sim \sigma^2$,
the vacuum $\sigma =0$ remains supersymmetric, while in all other vacua supersymmetry
is broken.

\section{\cpn model at small heterotic  deformations}
\label{hecpnsm}
\setcounter{equation}{0}

Now, we switch on the heterotic deformation which breaks \ntt\, supersymmetry down to \nzt\!.
In this section we will assume this deformation to be small limiting ourselves to the lowest
nontrivial order in the heterotic deformation. All preparatory work was carried out in Sect.~\ref{lnscptm}.
Therefore, here we can focus on the impact of the heterotic deformation {\em per se}.

To determine the effective action allowing us to explore the vacuum structure of the
heterotic
model, just as in Sect.~\ref{lnscptm}, we integrate over all but one given $ n^l $ field (and its superpartner $ \xi^l $).
One can always choose this fixed (unintegrated) field to be   $n^0\equiv n$.
Assuming $\sigma$ and $D$ to be constant background fields,
and  evaluating the determinants
one arrives at the following effective potential (see Eq.~(\ref{deffpp})):
\beqn
	V_\text{eff} & =& \int d^2x 
		\biggl\{  \left( iD ~+~ \bigl|\sqrt{2}\sigma -  m_0 \bigr|^2 \right) |n|^2 
	\nonumber\\[4mm]
	&-& 
	\frac{1}{4\pi}\, \sum_{i=1}^{N-1} \left( iD ~+~ \bigl|\sqrt{2}\sigma -  m^i \bigr|^2 \right)\,
		\ln\, \frac{ iD \,+\, \bigl| \sqrt{2}\sigma - m^i \bigr|^2} {\Lambda^2}
\label{Veff}
\\[4mm]
	&+&
	\frac{1}{4\pi}\, \sum_{i=1}^{N-1} \bigl|\sqrt{2}\sigma - m^i \bigr|^2\,
			\ln\, \frac{ \bigl| \sqrt{2}\sigma - m^i \bigr|^2 } { \Lambda^2 }
	~+~
	\frac{1}{4\pi}\, iD\, (N-1) 
	~+~
	\frac{N}{2\pi} \cdot u\, \bigl|\sigma\bigr|^2 \biggr\} \,,
	\nonumber\
\eeqn
where we have introduced a deformation parameter
 \beq
u ~~\equiv~~ \frac{8\pi}{N}\,|\omega|^2 \,.
\label{u}
\eeq	
Note that although $ |\omega|^2 $ grows as $ O(N) $ for large $N$, the parameter $u$ 
does not scale with $N$ and so is more appropriate for the r{o}le of an expansion parameter. 

The above expression for $ V_\text{eff} $ replicates Eq.~(\ref{Veff22}) except for the last term 
representing the heterotic deformation.
Now, to find the vacua, we must minimize the 
effective potential (\ref{Veff}) with respect to $n$, $D$ and $ \sigma $. 
The set of the vacuum equations is
\beqn
\label{eff1}
	&&
	|n|^2  -  \frac{1}{4\pi} \,\sum_{i=1}^{N-1}\, 
		{\rm ln}\, 
		\frac{i D + \bigl| \sqrt{2}\sigma -  m^i \bigr|^2}
							{\Lambda^2}   = 0\,, 
\\[3mm]				
	&&
	\left( i D + \bigl|\sqrt{2}\sigma - m_0 \bigr|^2 \right) n = 0\,, 
	\label{eff2}
	\\[3mm]
	&&
	\bigl(\sqrt{2}\sigma - {m_0} \bigr)|n|^2 
		-
	\frac{1}{4\pi} \sum_{i=1}^{N-1}\,
			\bigl(\sqrt{2}\sigma - m^i \bigr)\,
		{\rm ln}\, 
		\frac{i D + \bigl| \sqrt{2}\sigma \,-\,  m^i \bigr|^2}
		{\bigl| \sqrt{2}\sigma \,-\,  m^i \bigr|^2 }
		+ \frac{N}{4\pi} \cdot u\, \sqrt{2}\sigma = 0\,.
\label{eff3}
\nonumber\\
\eeqn
It is identical to the master set of Sect.~\ref{lnscptm}
with the exception of the last term in Eq.~(\ref{eff3}).
Equation (\ref{eff2}) is the same; hence we have the same two options:
 either
\beq
\label{higgsph}
	 iD + \bigl| \sqrt{2}\sigma - { m_0} \bigr|^2 = 0  
\eeq
or
\beq
\label{strongph}
	 n = 0 \,. 
\eeq
Since the deformation parameter is assumed to be small, we 
will solve these equations perturbatively, expanding in powers of  $ u $,
\beqn
	n &=& \nz  + u \cdot \no + \ldots\,,
	\nonumber
	 \\[3mm]
	iD &=& i\Dz  + u \cdot i\Do + \ldots\,, 
	\nonumber
	\\[3mm]
	\sigma &=& \sigz + u \cdot \sigo + \ldots\,.
\eeqn
	Here $ \nz $, $ \Dz $ and $\sigz$ constitute the solution of the \ntwot CP($N-1$) sigma model,
	in particular $ \Dz = 0 $ in both cases (\ref{higgsph}) and  
	(\ref{strongph}) corresponding to the Higgs and the strong-coupling regimes of the theory,
	respectively. We remind that the mass parameters are chosen according to (\ref{two}).

%
%
\subsection{The Higgs regime}
\label{subshr}

	The large-$N$ supersymmetric solution of the \ntwot CP($N-1$) sigma model
	in the Higgs phase is
given in Eqs.~(\ref{higgsvacN}) and (\ref{22higgscoupling}).
Expanding Eqs. (\ref{eff1}) -- (\ref{eff3}) to the first order in $u$, we calculate
\beqn
	i \Dz 
	&=&
	 0\,,                    \qquad\quad      i\Do   = 0\,,  
			\qquad iD^{(2)} =  -\,|\sqrt{2}\sigo|^2\,, 
			\nonumber\\[3mm]
	\sqrt{2}\sigz &=& m,  \qquad       
	\sqrt{2}\sigo  = -\,\frac{N}{4\pi}\,\frac{m}{|\nz|^2}\,, 
	 \nonumber\\[3mm]
	|\nz|^2 
	&=&
	2\beta^{(0)}_{\rm ren}\,,              \qquad   
	\no =  -\,\frac{2\,m}{\bnz\,|\nz|^2}\,\frac{N}{32\pi^2}\,\sum_{i=1}^{N-1} \frac{1}{m - m^i}\,.
			\label{higgseqp}
			\eeqn
	With masses from (\ref{two}) we then obtain
\beq
	\sum_{i=1}^{N-1} \frac{1}{m - m^i} ~=~ \frac{N-1}{2m} ~=~ \frac{N}{2m} ~+~ O(1)\,.
	\label{higgseqpp}
\eeq
Using this, we simplify the solution (\ref{higgseqp}),
\beqn
	\sqrt{2}\sigma 
	&=&
	 m \lgr 1 ~-~ \frac{u/2}{{\rm ln}\, m/\Lambda} \rgr ~+~ \ldots \,,
	\nonumber\\[2mm]
	i D  
	&=&
	-\, m^2 \left( \frac{u/2}{{\rm ln}\, m/\Lambda} \right)^2  ~+~ \ldots\,,
	 \label{finhi}
	\\[2mm]
%
	n 
	&=&
	\sqrt{2\beta^{(0)}_{\rm ren}} \lgr 
			1 ~-~ \frac{ u/8 } { ( {\rm ln}\, m/\Lambda )^2 } \rgr ~+~ \ldots\;.
	\nonumber
\eeqn
This is in the Higgs phase, where
$$
2\beta^{(0)}_{\rm ren} ~=~ \frac{N}{2\pi}\,{\rm ln}\left(\frac{m}{\Lambda}\right)\,.
$$
Eqs.~\eqref{finhi} agree with numerical calculations of the solution of the 
vacuum equations in the Higgs phase. 

%
%
\subsection{Strong coupling}
\label{subsestrco}

Our starting point is the zeroth order in $ u $ solution (Sect.~\ref{lnscptm}),
\beq
	\nz ~=~ 0\,, \quad
	i \Dz ~=~ 0\,, \quad
	\sqrt{2}\sigz ~=~ \wt{\Lambda}\cdot e^{i\frac{2\pi l}{N}}\,,  
	\label{zosol}
	\eeq
	where
\beq
	\wt{\Lambda} ~=~ \sqrt[N]{ \Lambda^N + m^N } ~=~ \Lambda\lgr 1 ~+~ O\left(e^{-N}\right) \rgr\,\, 
	{\rm at}\,\,\, N ~\to~ \infty\,.
	\label{tilla}
\eeq
	At strong coupling $n$ vanishes exactly, not only in the zeroth order in $u$.
Omitting the details, the first order solution to the vacuum equations (\ref{eff1}), (\ref{eff3}) 
is given by (in conjunction with $n=0$)
\beqn
&&
 \Dz  ~~=~~ 0\,,\qquad
	i \Do ~~=~~ \frac{ \sqrt{2}\sigz } 
		    {\frac{1}{N} \sum_{i=1}^{N-1} 
		                 \frac{1}{ \sqrt{2}\bsigz - \ov{m}{}^i } } \,,
\label{strongeq}\\[3mm]
&&
	\sqrt{2}\sigo \, \frac{1}{N} \sum_{i=1}^{N-1} \frac{1}
			{ \sqrt{2}\sigz - m^i }
		~+~ \text{h.c.} ~~=~~
	-\, \sqrt{2}\sigz  \,\,
	\frac{ \sum_{i=1}^{N-1} 
			\frac{1}{ | \sqrt{2}\sigz - m^i |^2 } }
	  {  \sum_{i=1}^{N-1} \frac{1}{ \sqrt{2}\bsigz - \ov{m}{}^i } }\,.
\notag
\eeqn
We use the following relations to simplify the above formulas when masses are set as in (\ref{two}):
\begin{align}
	\frac{1}{N}\, \sum_{k=0}^{N-1} \frac{1} { 1 \,-\, \alpha\,e^{\frac{2\pi i k}{N}} } &~~=~~
		\frac{1}{1 \,-\, \alpha^N} ~~\simeq~~ 1\,,
	\nonumber\\[2mm]
	\frac{1}{N}\, \sum_{k=0}^{N-1} \frac{1} { (1 + \alpha^2) - 2\,\alpha \cos \frac{2 \pi k}{N} }
		&~~=~~
	\frac{1} {1 \,-\, \alpha^2} \,\,\, \frac{1 \,+\, \alpha^N}{ 1 \,-\, \alpha^N} 
		~~\simeq~~
	\frac{1} {1 \,-\, \alpha^2} \,.
	\label{vspfor}
\end{align}
	This enables us to present the results for $m\ll\Lambda$ in the following quite simple form:
\beqn
\label{finstr}
	n 
	&=&  0 \,,\qquad i D   ~=~  u\, \Lambda^2 ~+~ \ldots \,, 
	\nonumber\\[3mm]
	\sqrt{2}\sigma & =& \Lambda \, e^{\frac{2\pi i l}{N}} \lgr
			1 ~-~ \frac{u}{2}\, 
				\frac{\Lambda^2}{\Lambda^2 \,-\, m^2}\,\rgr ~+~ \ldots.
\eeqn
	We complement these formulas for the strong coupling phase by finding the approximate solutions
	now as expansions in $m^2$ parameter, assuming $m$ to be small.
	We obtain
\begin{align}
\notag
	\sqrt{2}\sigma & ~=~ e^{\frac{2\pi i l}{N}} \Lambda \lgr 
			   e^{-u/2} ~-~ \frac{m^2}{\Lambda^2}\, {\rm sh}\,u/2 \rgr ~+~ \ldots\;,
	\\
\label{smstr}
	iD & ~=~ \Lambda^2 \left( 1 ~-~ e^{-u} \right) ~+~ O\left(\frac{m^4}{\Lambda^4}\right)\, ,
\end{align}
	where $u$ does not need to be (too) small anymore.

	Just a brief look at the Higgs phase solution (\ref{finhi}) and the strong coupling phase solutions
	(\ref{finstr}) and (\ref{smstr}) reveals, that these expansions blow up 
	when one approaches $ m ~\approx~ \Lambda $! 
	While the exact solutions are expected to be finite for all $m$, our approximations cannot be trusted
	at $ m ~=~ \Lambda $.
	This is the first sign that something is going on at these values of masses.
	As we will later see from the large-$u$ solutions, as well as from the numerical solution of the 
	vacuum equations, the theory experiences a double phase transition as $m$ goes from the area
	below $ \Lambda $ towards $ m \gg \Lambda $.

\section{ Heterotic \cpn model at large  deformations}
\label{hetdefld}
\setcounter{equation}{0}

Now it is time to study equations (\ref{eff1}) -- (\ref{eff3}) in the opposite limit of large values of 
the deformation parameter $u\gg 1$. We will see that our theory has three distinct phases 
separated by two phase transitions:

\vspace{1mm}

 (i) Strong coupling phase  with the broken $Z_N$
symmetry at small $m$;

(ii)  Coulomb/confining $Z_N$-symmetric
phase at intermediate $m$ (the coupling constant is strong in this phase as in the case (i)); 

(iii) Higgs phase
at large $m$ where the $Z_N$ symmetry is again broken.

\vspace{1mm}

As previously,  we assume that mass parameters are chosen in accordance with
(\ref{two}).

\subsection{ Strong coupling phase with broken \boldmath{$Z_N$}}
\label{scpwbz}

This phase occurs at very small masses, namely,
\beq
m\le \Lambda\,e^{-u/2}\,,\qquad u\gg 1\,.
\label{scphmass}
\eeq
In this phase we have 
\beq
|n|=0,\qquad 2\bren=\frac{1}{4\pi}\, 
\sum_{i=1}^{N-1}\ln\, {\frac{iD +|\sqrt{2}\sigma-m_i|^2}{\Lambda^2}} =0\, .
\label{scphn}
\eeq
As we will see momentarily,  $\sigma$ is exponentially small in this phase. Masses are also
small. 
Then the second equation in (\ref{scphn}) gives
\beq
iD ~~\approx~~ \Lambda^2 \,.
\label{scphD}
\eeq
With this value of $iD$ we can rewrite Eq.~(\ref{eff3}) in the form
\beq
 \sum_{i=1}^{N-1}\left(\sqrt{2}\sigma-m_i\right)\ln\, {\frac{\Lambda^2}{\left|\sqrt{2}\sigma-m_i\right|^2}}
=N\left(\sqrt{2}\sigma\right)u\,.
\label{scpheq3}
\eeq
The following trick is very convenient for solving this equation.

\vspace{2mm}

Let us consider an auxiliary problem from static electrodynamics in two dimensions.
Assume we have $N$ equal ``electric charges" evenly distributed over the circle
depicted in Fig.~\ref{circ}. In the limit of large $N$ one can consider this distribution to be continuous
(and homogeneous). The task is to find the electrostatic potential at the point $x$ on the plane.

\begin{figure}
\epsfxsize=7cm
\centerline{\input{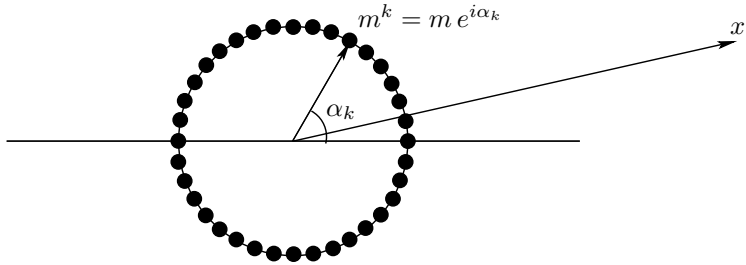}}
\caption{\small Electrostatic analog problem. The circle of radius $m$ (see Eq.~(\ref{two}))
is homogeneously populated by ``electric charges," namely,
$\alpha_k = 2\pi \, k/N$ where $k= 0,1,2, ..., N-1$.
We then must calculate the electrostatic potential at the
point $x$.}
\label{circ}
\end{figure}
It is not difficult to calculate  the potential of a charged circle
of radius $m$  centered at the origin  in  two-dimensional electrostatics.
Representing $x$ by a complex  number we get (in the large-$N$ limit) 
\beq
\frac{1}{N}\sum_{i=0}^{N-1}\ln\, {|x-m_i|^2}
=\left\{
\begin{array}{cc}
\ln\,{|x|^2},\;\;\;\;\; |x|>m\\[2mm]
\ln\,{m^2},\;\;\;\;\;\; |x|<m
\end{array}\,.
\right.
\label{chargedcircle}
\eeq
Now, to obtain the left-hand side of (\ref{scpheq3}) we must integrate (\ref{chargedcircle})
over $x$ and then substitute $x=\sqrt{2}\sigma$. In this way we arrive at
\beq
 \frac{1}{N}\sum_{i=0}^{N-1}\left(\sqrt{2}\sigma-m_i\right)\ln\, {\frac{\Lambda^2}{|\sqrt{2}\sigma-m_i|^2}}
=\left\{
\begin{array}{cc}
\sqrt{2}\sigma\,\ln\, {\frac{\Lambda^2}{|\sqrt{2}\,\sigma|^2}}-\frac{m^2}{\sqrt{2}\,\bar\sigma}\,,\quad |\sqrt{2}\sigma|>m\\[3mm]
\sqrt{2}\sigma\,\left(\ln\, {\frac{\Lambda^2}{m^2}}-1\right),\qquad |\sqrt{2}\sigma|<m
\end{array}
\right. .
\label{usefulformula}
\eeq
Outside the circle the potential is the same as that of the unit charge at the origin.
Inside the circle the potential is constant.

Substituting Eq.~(\ref{usefulformula}) in (\ref{scpheq3}) at  $m<|\sqrt{2}\sigma|$ 
(i.e. outside the circle) we get
\beq
\sqrt{2}\langle \sigma\rangle=e^{\frac{2\pi i}{N}k}\;\Lambda\,e^{-u/2},\qquad k=0,...,(N-1).
\label{scphsigma}
\eeq
The vacuum value of $\sigma$ is exponentially small at large $u$. The bound $m<|\sqrt{2}\sigma|$
translates into the condition (\ref{scphmass}) for  $m$. 

We see that we have $N$ degenerate vacua in this phase. The  chiral $Z_{2N}$ symmetry is broken
down to $Z_2$, the order parameter is $\langle \sigma\rangle$. Moreover, the absolute 
value of $\sigma$ in these vacua does not depend on $m$.
In fact, this solution coincides with the one obtained in \cite{SYhet} for $m=0$. This phase
is quite similar to the strong coupling phase of  the \ntwot \cpn model, see (\ref{22sigmaapp}).
The difference is that the absolute value of $\sigma$ depends now on $u$ and becomes exponentially
small in the limit $u\gg 1$.

The vacuum energy is positive (see Eq. (\ref{scphD})) --  supersymmetry is broken.
We will present a plot of the vacuum energy as a function of $m$ below, in Sect.~\ref{subscoulco}.

\subsection{Coulomb/confining phase}
\label{subscoulco}

Now we increase $m$ above the bound (\ref{scphmass}). From (\ref{usefulformula}) we see that the 
exponentially small solution to Eq.~(\ref{scpheq3}) no longer exist. The only solution is
\beq
\langle \sigma\rangle=0\,.
\label{confsigma}
\eeq
In addition, Eq.~(\ref{scphn}) implies
\beq
|n|=0,\qquad iD= \Lambda^2-m^2\,.
\label{confnD}
\eeq
This solution describes a single $Z_N$ symmetric vacuum. All other vacua are lifted
and become quasivacua (metastable at large $N$). This phase is quite similar to the Coulomb/confining phase
of nonsupersymmetric \cpn model without twisted masses \cite{W79}. The presence 
of small splittings between quasivacua produces a linear rising confining potential between kinks
that interpolate between, say, the true vacuum and the lowest quasivacuum \cite{GSY05},
see also the review \cite{SYrev}.
As was already mentioned, this linear potential was  interpreted, long time ago \cite{Coleman,W79}, as the
 Coulomb interaction,
 see the next section for a more detailed discussion.

As soon as we have a phase with 
the broken $Z_N$ symmetry at small $m$,
 and the $Z_N$-symmetric phase at
intermediate $m$ the theory experiences
a phase transition that separates these phases. As a rule, 
one does not 
have phase transitions in supersymmetric theories. However, in the model at hand  supersymmetry is
badly broken (in fact, it is broken already at the classical level \cite{BSY3}); therefore, the emergence of 
a phase transition is not too surprising. 

We can calculate the vacuum energy  explicitly to see  the degree of supersymmetry breaking.
Substituting (\ref{confsigma}) and (\ref{confnD}) in the effective potential (\ref{Veff})
we get
\beq
E_{\rm vac}^{\rm Coulomb}~=~\frac{N}{4\pi}\left[ \Lambda^2~-~m^2~+~m^2\ln {\frac{m^2}{\Lambda^2}}\right].
\label{confEvac}
\eeq
The behavior of the vacuum energy density $E_{\rm vac}$ vs. $m$ is shown  in Fig.~\ref{figvacE}.

\begin{figure}
\epsfxsize=10cm
\centerline{\input{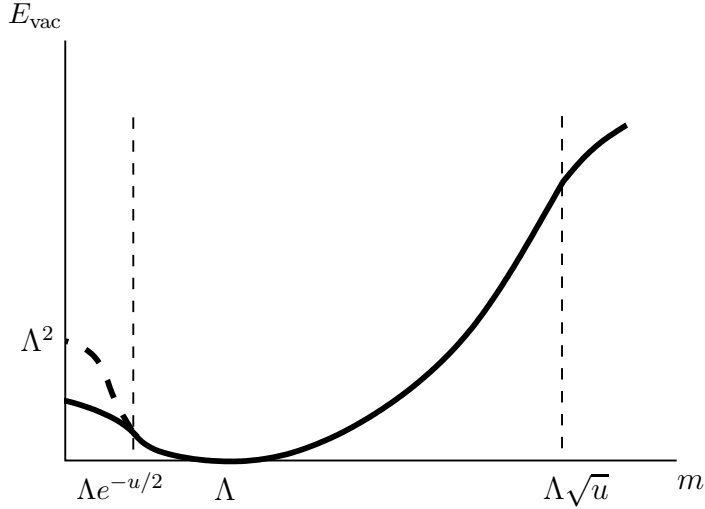}}
\caption{\small Vacuum energy density vs. $m$. The dashed line shows the behavior of the
energy density \eqref{confEvac} extrapolated into the strong coupling region.} 
\label{figvacE}
\end{figure}

$E_{\rm vac}$ is positive  at generic values of $m$, as it should be in the
 case of the spontaneous breaking of supersymmetry.
Observe, however, that the vacuum energy density {\em vanishes}
at $m=\Lambda$. This is a signal of \ntwoo supersymmetry restoration. To check that this
is indeed the case -- supersymmetry is dynamically restored at  $m=\Lambda$ -- we can
compare the masses of the bosons $n^i$ and their fermion superpartners $\xi^i$. 
From (\ref{bee31}) we see that the difference of 
their masses reduces to $iD$. Now, Eq.~(\ref{confnD}) shows that $iD$ vanishes exactly at $m=\Lambda$.

This is a remarkable phenomenon: while \ntwoo supersymmetry is broken at the classical level
at $m=\Lambda$, 
it gets restored at the quantum level at this particular point in the parameter space. 
This  observation is implicit
 in \cite{Tonghetdyn} where a Veneziano--Yankielowicz-type (VY-type)
 superpotential  \cite{VYan} for  \ntwot \cpn model (see \cite{AdDVecSal,ChVa,W93}) was
 extrapolated to   the \ntwoo  case.

We pause here to make an explanatory remark regarding Fig.~\ref{figvacE} and Eq.~(\ref{confEvac}).
The plot of $E_{\rm vac}^{\rm Coulomb}$ is presented in this figure assuming the parameter $m$
to be real (we follow this assumption in the bulk of the paper). In fact, $m$ can be viewed as a complex parameter,
the phase of $m$ being interpreted as a $\theta$ angle. A straightforward examination shows
that for the complex values of $m$ Eq.~(\ref{confEvac}) must be replaced by
$$
E_{\rm vac}^{\rm Coulomb}~=~\frac{N}{4\pi}\left[ \Lambda^2~-|m|^2~+|m|^2\ln {\frac{|m|^2}{\Lambda^2}}\right].
$$
This means that the vacuum is supersymmetric (i.e. $E_{\rm vac}^{\rm Coulomb}=0$) on the curve
$|m|^2=\Lambda^2$.


Now we turn our attention to what happens with $\sigma $ at the strong/Coulomb
phase transition point.
More detail on that will be given in Section \ref{subsnum} with the help
of numerical calculations, however, at large $u$ the behavior of $\sigma$ can be analyzed 
just by inspecting Eq.~(\ref{scpheq3}). 
To solve this equation we can evaluate the sum in it using Eq.~(\ref{usefulformula}).
In place of the massive quantities $ \sigma $, $\Lambda $ and $ m $ it is convenient to 
introduce dimensionless variables
\beq
	\cs ~~=~~ \frac{\sqrt{2} \sigma}{\Lambda}\,, \qquad\qquad
        \mu ~~=~~ m/\Lambda\,.
\label{csdef}
\eeq
Then Eq. (\ref{scpheq3}) turns into
\beq
\label{mu-s}
	\mu^2  ~~=~~ - \cs^2 \lgr u  ~+~ \ln \cs^2 \rgr .
\eeq
Instead of solving for $ \cs $ one can use (\ref{mu-s}) as a solution for $ m $ with respect 
to $ \sigma $.
\begin{figure}
\centerline{\resizebox{11cm}{!}{\input{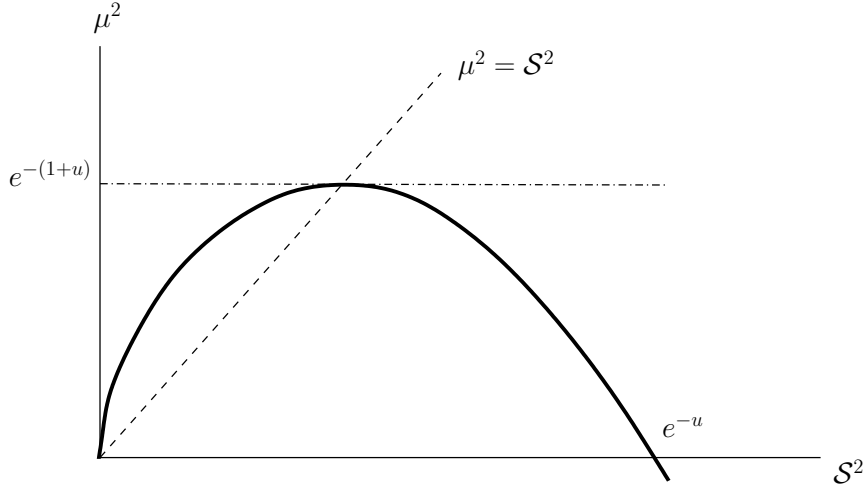}}}
\caption{\small Dependence of $m^2$ versus $\sigma^2$. The rescaled variables as in text are
$\mu = m/\Lambda$ and $\cs = \sqrt{2}\sigma/\Lambda$. }
\label{fig:mus}
\end{figure}
Figure~\ref{fig:mus} illustrates the dependence (\ref{mu-s}). 
We now treat this graph as the dependence of $ \sigma $ on  $m$.
In particular, for zero masses, $ \sqrt{2}\sigma = \Lambda e^{-u/2} $, as it should be.
Following from right to left, as the mass grows, $ \sigma $ decreases, and at some point 
the curve bends downward. 
This happens at $ \mu^2 = \cs^2 = e^{-(1+u)}$, as can be seen from Eq.~(\ref{mu-s}). 
At this point the derivative $ \p\sigma/\p m $ becomes infinite, which is indicative of a
phase transition. 
The behavior of $ \sigma $ near this point is circle-like,
\beq
\label{scirc}
	\sqrt{2}\sigma ~~\simeq~~ \Lambda\, e^{-\, \frac{1+u}{2}}  ~+~ 
		\frac{1}{2}\,\Lambda\, \sqrt{e^{- (u+1)} ~-~ m^2/\Lambda^2 } \,.
\eeq
Equations (\ref{mu-s}) and  (\ref{scirc}) together with Fig.~\ref{fig:mus} are approximate, but qualitatively
they very closely demonstrate what happens to $ \sigma $ in reality. 
Namely, $ \sigma $ monotonically decreases with increase of mass, 
until it experiences a vertical bend, at which point
it drops down to zero, the fact that cannot be seen from Eq.~(\ref{mu-s}). 
Rather, this is seen in the Coulomb phase, and reproduced with numerical solution in Section~\ref{subsnum}.
Despite the fact that apparently $ \sigma $ experiences a drop, the energy density does not
(see Fig.~\ref{fig:numEm}), and therefore the phase transition is of the second order. 

\subsection{ Higgs phase}
\label{subshiggph}

The Higgs phase occurs in the model 
under consideration at large $m$. Below we will show that the model is in the Higgs phase at
\beq
m ~~>~~ \sqrt{u}\Lambda\,,\qquad {\rm if} \,\,u~~\gg~~ 1\,.
\label{Hphmass}
\eeq
In this phase $|n|$ develops a VEV. From Eq.~(\ref{eff2}) we see that
\beq
iD ~=~ -\bigl|\sqrt{2}\sigma -m_0\bigr|^2\, .
\label{HphD}
\eeq

To begin with, let us examine  Eqs.~(\ref{eff1}) -- (\ref{eff3}) 
far to the right from the boundary (\ref{Hphmass}), i.e. 
at $m\gg \sqrt{u}\Lambda$. 
In this regime we can drop the second logarithmic term in (\ref{eff3}).
This will be confirmed shortly.
 The first
term is much larger because it is   proportional to $\bren$ which is large in the quasiclassical region
(see Eqs. (\ref{22eff1}) and  (\ref{higgsvac})). Then Eq.~(\ref{eff3}) reduces to
\beq
(\sqrt{2}\sigma ~-~m_0)\,2\bren ~+~ \frac{N}{4\pi}\,u\,\sqrt{2}\sigma ~~=~~0 \,,
\eeq
implying, at large $u$  
\beq
\sqrt{2}\sigma ~~=~~ \left(\frac{8\pi}{N}\bren\right)\,\frac{m_{{0}}}{u},
\label{Hphsigma}
\eeq
where we take into account 
 that $|\sigma |\ll m$, the fact justified {\em a posteriori}. Equation (\ref{Hphsigma}) applies to the 
 $k=0$ vacuum. It is obvious that the solution for other $N-1$ vacua can be obtained
 from (\ref{Hphsigma})  by replacing $m_0\to m_{i_0}$ where $i_0=1,...,(N-1)$.

Thus, we have $N$ degenerate vacua again. In each of them $|\sigma |$ is small ($\sim m/u$) but
nonvanishing. The $Z_{2N}$ chiral symmetry is again broken down to $Z_2$. Clearly, the 
Higgs phase is separated form the Coulomb/confining phase (where $Z_{2N}$ is unbroken) by
a phase transition. 

To get the vacuum expectation value of $n^{0}$ we must analyze the logarithms in Eq.~(\ref{eff1}) and (\ref{eff3})
with a better accuracy: $\sigma$ in the numerators cannot be neglected. We must keep the terms linear in $\sigma$.
Since the solution for $\sigma$ is real, see (\ref{Hphsigma}), 
we can rewrite the logarithm in (\ref{eff1}) as follows:
\beqn
&&
{\rm ln}\left( i D + \bigl| \sqrt{2}\sigma -  m^i \bigr|^2 \right)
~=~
{\rm ln}\left(2\sqrt{2} \sigma\,{\rm Re} (m_0 - m_i)\right)
\nonumber\\[3mm]
&&=~
{\rm ln}\left[4\sqrt{2} \sigma\, m\, \sin^2\left(\frac{\alpha_k}{2}\right)\right],\qquad \alpha_k =\frac{2\pi\,k}{N}\,,
\quad k= 1,..., N-1\,.
\label{hujone}
\eeqn
where $\alpha_k$ is the phase of $m_k$, see Fig.~\ref{circ}. On the other hand, Eq.~(\ref{22higgscoupling})
can be presented in the form
\beq
\frac{1}{4\pi}\, \sum_{k=1}^{N-1} \, {\rm ln}\left[4\, m^2\, \sin^2\left(\frac{\alpha_k}{2}\right)\right]
= \frac{N}{4\pi}\,\ln m^2\,.
\label{hujtwo}
\eeq
Thus, we conclude that
\beqn
|n|^2 
&=&
2\bren ~=~ \frac{N}{4\pi}\, \ln \frac{\sqrt{2}\sigma\, m}{\Lambda^2} 
\nonumber\\[3mm]
&\sim&
  \frac{N}{4\pi}\, \ln \frac{ m^2}{u\,\Lambda^2} 
\label{hujthree}
\eeqn
in each of the $N$ vacua in the Higgs phase. Here the last (rather rough) estimate follows from (\ref{Hphsigma}).

Our next task is to get an equation for $\bren$ ({\em en route}, we will relax 
the constraint $m\gg\sqrt{u}\Lambda$). To this end we must examine Eq.~(\ref{eff3}),
including the logarithm into consideration. We will expand the numerator neglecting $O(\sigma^2)$ terms,
while in the denominator we can set $\sigma =0$ right away. Then the summation in 
(\ref{eff3}) can be readily performed
using the formula
\beq
\frac{1}{N}\,\sum_{k=0}^{N-1}\, \left(m_0-m_k\right)\ln\left[4\,m^2\,\sin^2\left(\frac{\alpha_k}{2}\right)\right]
=m\left(\ln m^2+1\right),
\label{hujfour}
\eeq
which follows, in turn, from Eq.~(\ref{usefulformula}). As a result, we arrive at
\beq
\sqrt{2}\sigma\,u = m\left(\frac{8\pi}{N}\,\bren +1\right).
\label{hujfive}
\eeq
The only approximation here is $u\gg 1$, plus, of course, Eq.~(\ref{Hphmass}). Combining Eqs.~(\ref{hujfive})
and (\ref{hujthree}) we obtain the following relation for $\bren$:
\beq
\frac{8\pi}{N}\bren- \ln{\left(\frac{8\pi}{N}\,\bren+1 \right)}=\ln{\frac{m^2}{u\Lambda^2}}  \,.
\label{Hphbetaeq}
\eeq
Strictly speaking, Eq.~(\ref{Hphbetaeq}) has two solutions at large $m$, deep inside the Higgs domain. 
The smaller solution corresponds to  negative $\bren$. Since $|n|^2=2\bren$ is positively defined we keep only
 the larger one. At $m \gg\sqrt{u}\Lambda\;$ $\bren$ is large and is given 
(with the logarithmic accuracy) by the last estimate
in Eq.~(\ref{hujthree}).
As we reduce $m$, at $m = \sqrt{u}\Lambda$, two solutions of (\ref{Hphbetaeq}) coalesce. 
At smaller $m$ they become complex. Thus
$m = \sqrt{u}\Lambda$ is indeed the phase transition point to the Coulomb/confining phase.
At this point $\bren=0$, which coincides with its value in the Coulomb/confining phase, see
(\ref{confnD}). Thus, $\bren$ is continuous at the point of the phase transition, while its
derivative over $m$ is discontinuous.

Calculating the vacuum energy in this phase we get
\beq
E_{\rm vac}^{\rm Higgs} ~~=~~ \frac{N}{4\pi}\lgr m^2\ln {\frac{m^2}{\Lambda^2}} ~+~ \Lambda^2 ~-~ m^2 
                                                         ~+~ O \left( \frac{m^2}{u^2} \ln \frac{m^2}{\Lambda^2} \right)
					\rgr.
\label{HiggsEvac}
\eeq
The vacuum energy in the Higgs phase is non-zero so the \ntwoo supersymmetry is broken. It was 
observed earlier in \cite{BSY3} on the classical level, see also (\ref{ferbosmasssplit}).
The vacuum energy density in all phases is displayed in Fig.~\ref{figvacE}. 
We can see that the expression for energy \eqref{HiggsEvac} at large $u$ coincides with Eq.~\eqref{confEvac},
which signifies that the phase transition is of the second order.
In Section~\ref{subsnumhiggs} we will analyze the transition in more detail.

%
%
\subsection{Evaluation at arbitrary magnitude of deformation} 
\label{subsnum}

In this subsection we will grasp a picture of what happens to the order parameters in phase 
transition regions when the deformation $ u $ is not necessarily large or small.
Although we will be able to acquire an insight into the solution of the vacuum equations
in the Higgs phase analytically, we will need to revoke numerical calculations to 
examine the energy density and the coupling constant. 
In the Strong phase, even less can be done analytically, and we will resort to numerical
methods almost entirely. 

\subsubsection{Strong/Coulomb phase transition}
\label{subsnumstrong}

\begin{figure}
\epsfxsize=10cm
\centerline{\epsfbox{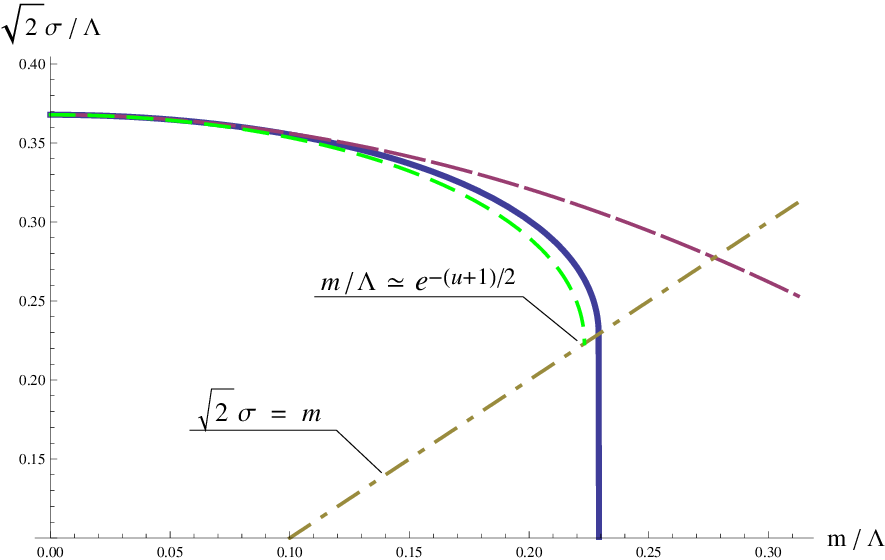}}
\caption{\small The dependence of $\sigma$ on $m$. The solid curve shows the numerical solution
of the equations \eqref{eff1} and \eqref{eff3} in the strong coupling phase ($|n| = 0 $) for $u = 2$.
The upper dashed curve shows the small-$m$ limit \eqref{smstr}.
The lower dashed curve refers to the large-$u$ solution obtained as a numerical evaluation of
Eq.~\eqref{mu-s}.
The dash-dot $ \sqrt{2}\sigma = m $ line is given for reference.
To the right of the drop, $ \sigma $ is identically zero.}
\label{fig:numsigm}
\end{figure}
To see what happens at the strong-to-Coulomb phase transition we solve the vacuum equations 
(\ref{eff1}) -- (\ref{eff3}) numerically, with $ n ~=~ 0 $.
Figure~\ref{fig:numsigm} shows the dependence of $ \sigma $ on $ m $ in the strong
coupling phase. 
The main graph is compared to the small-$m$ solution (\ref{smstr}) and the large-$u$
dependence (\ref{mu-s}) (still, solved numerically). 
One convinces that the large-$u$ solution given by Eqs.~(\ref{scpheq3}) and (\ref{mu-s})
describes the true solution really well. 
The curve for $\sqrt{2}\sigma$ monotonically decreases from the value $\Lambda e^{-u/2}$, 
until it meets the line $ \sqrt{2}\sigma = m $,
at which point $ \sigma $ turns down $ 90^o $ and drops vertically to zero. 
This is the strong-Coulomb phase transition point, which is approximately located at
$ m = \sqrt{2}\sigma \simeq e^{-(1+u)/2} $.
To the right of the phase transition point $ \sigma $ is identically zero. 

Figure~\ref{fig:numEm} shows the energy density in the strong coupling and Coulomb/confining phase. 
The curve clearly displays that the energy does not experience any jumps at the phase transition 
point $ m \simeq e^{-(1+u)/2} $.
That is, the phase transition is at most of the second order.
The curve, however, does apparently experience a break of incline at that point. 
To the right of the phase transition, the numerical curve exactly overlays the Coulomb phase
energy density Eq.~(\ref{confEvac}).
\begin{figure}
\epsfxsize=10cm
\centerline{\epsfbox{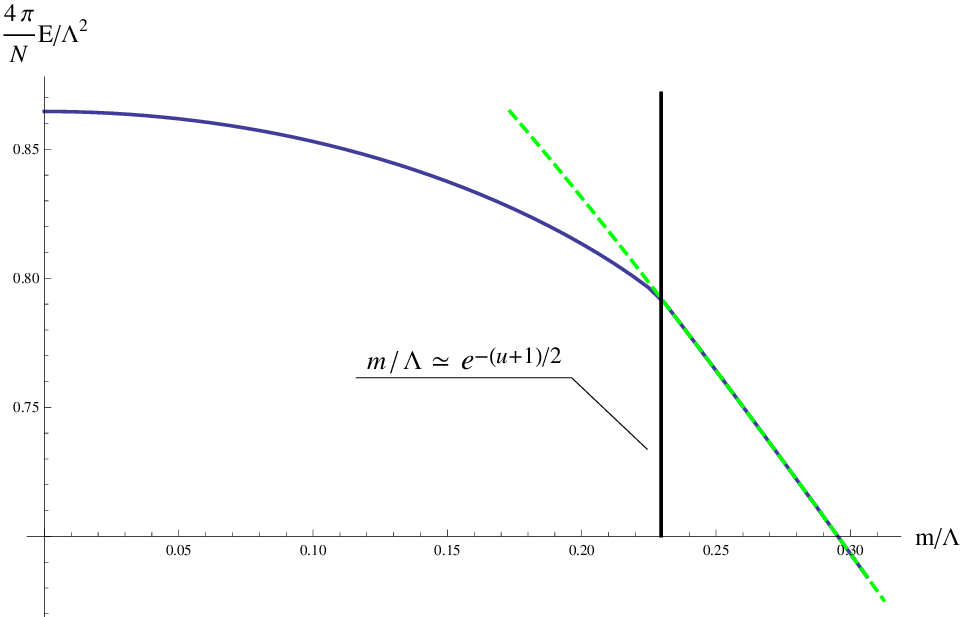}}
\caption{\small Energy density versus $m$ in the strong coupling and Coulomb phase for $u=2$. 
The dashed curve displays the energy density of the Coulomb/confining phase, Eq.~\eqref{confEvac}.}
\label{fig:numEm}
\end{figure}

We further clarify the position of the phase boundary numerically, by solving
the vacuum equations with the condition $ m = \sqrt{2} \sigma $. 
The phase boundary happens to coincide exactly with the left branch of the solution of 
the equation
\[
	u  ~~=~~ \mu_*^2 ~-~ \ln \mu_*^2 ~-~ 1\,,
\]
which will arise in the next subsection when we will be analyzing the Higgs phase transition.
At that point the full phase picture of the theory will be explicated. 

\subsubsection{Higgs phase}
\label{subsnumhiggs}

We revoke the dimensionless variables $ \cs $ and $ \mu $ introduced in Eq.~\eqref{csdef}.
The system of vacuum equations \eqref{eff1}-\eqref{eff3}, written in terms of these variables, becomes
\begin{align}
\notag
	\frac{1}{N}\,\sum_{k=1}^{N-1}  \left\{\,  (\mu^k - \mu)\, 
			                          \ln \lgr |\cs - \mu^k |^2  \,-\, |\cs - \mu |^2 \rgr 
                         			   ~+~ (\cs - \mu^k)\, \ln |\cs - \mu^k | ^2 \,\right\} ~+ \\
                              ~+~ u\cdot \cs ~~=~~ 0\,.
\end{align}
Here $ \mu_k ~=~ \mu\, e^{i2\pi k / N} $.
Evaluating the sum, one arrives at an algebraic equation
\beq
\label{higgs_vaceq}
	\left(\, 1 \,+\, u \,+\, \ln \mu^2 \,\right)\, \cs  ~~=~~ \mu \lgr 1 \,+\, \ln(\mu \cs) \rgr .
\eeq
We can solve Eq.~\eqref{higgs_vaceq} numerically. 
The result is summarized in Fig.~\ref{fig:numsignm} which shows the dependence of $ \cs $ and $ |n|^2 $ on $\mu$.
The latter dependence can be inferred from Eq.~\eqref{hujthree},
\beq
\label{higgs_numn}
	\frac{4\pi}{N}\, |n|^2 ~~=~~ \ln \frac{\sqrt{2}\sigma\, m}{\Lambda^2}
		~~=~~ \ln \mu\cs \,.
\eeq 

\begin{figure}
\epsfxsize=11cm
\centerline{\epsfbox{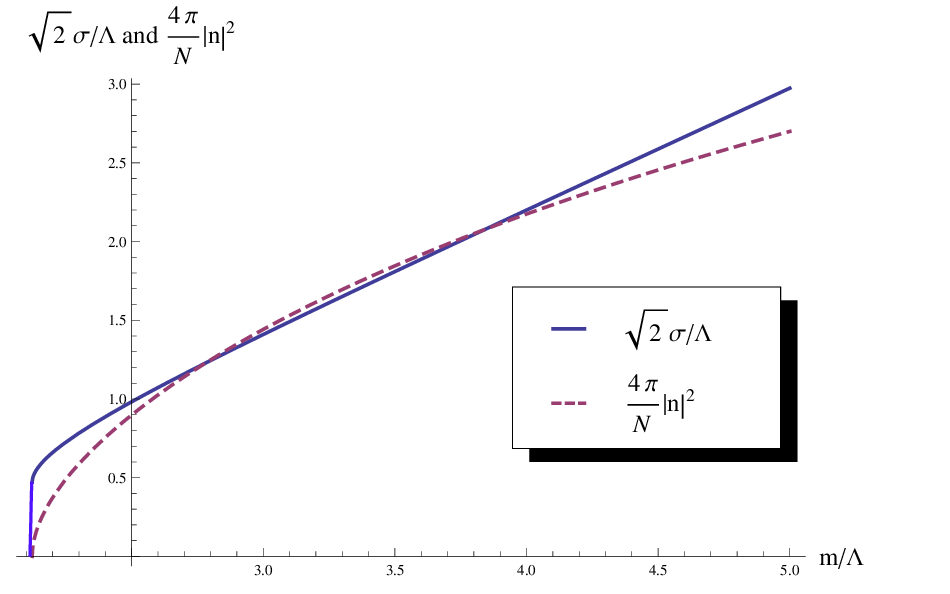}}
\caption{\small Dependence of $\sigma$ and coupling constant $|n|^2$ on $m$ in the Higgs phase for $u = 2$.
The former dependence is obtained from Eq.~\eqref{higgs_vaceq}, while the latter is then reconstructed from
Eq.~\eqref{higgs_numn}.}
\label{fig:numsignm}
\end{figure}
Vanishing of $ |n|^2 $ at certain $ \mu_* = m_*/\Lambda $ delineates the Higgs and the Coulomb/confining 
phases.
At that point, $\sigma$ experiences a vertical slope. 
To the left of the phase transition point, $ |n|^2 $ becomes negative --- 
the analysis of Eq.~\eqref{higgs_vaceq} is not valid.
Strictly speaking, Higgs phase vacuum equations become invalid as soon as $|n|^2$ reaches zero,
one needs to deal with the Coulomb phase. 

Demanding the derivative $ \p\cs/\p\mu $ in Eq.~\eqref{higgs_vaceq} to be infinite, 
one arrives at an equation governing the phase boundary:
\beq
\label{higgs_phbnd}
	\mu_*^2 ~-~ \ln \mu_*^2 ~~=~~ 1 ~+~ u\,,
\eeq
together with a useful relation
\beq
\label{higgs_cs}
	\mu_* ~~=~~ \frac{1}{\cs_*}\,.
\eeq
The first equation has two solutions, the corresponding curve shown on Fig.~\ref{fig:higgsborder}.
\begin{figure}
\epsfxsize=11cm
\centerline{\includegraphics[width=11.5cm,keepaspectratio]{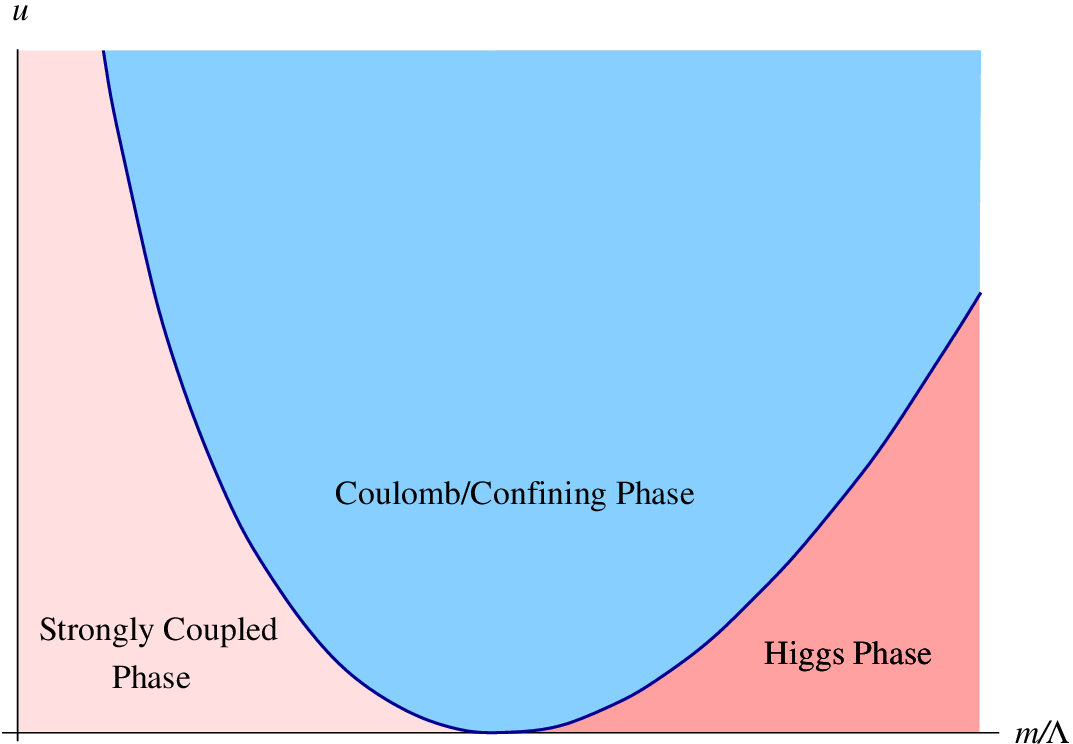}}
\caption{\small 
The phase diagram of the twisted-mass deformed heterotic CP($N-1$) theory.
Variable $u$ denotes the amount of deformation, $ u = \frac{8\pi}{N} |\omega|^2 $.
The phase transition boundaries are determined by Eq.~\eqref{higgs_phbnd}.
}
\label{fig:higgsborder}
\end{figure}
The larger solution determines the Coulomb/confining-Higgs phase boundary. Its asymptotics at large $u$,
$ \mu^2 \gg \ln \mu^2 $, is given by ({\it cf.} \eqref{Hphmass})
\beq
	\mu_{*H} ~~\simeq~~ \sqrt{ 1 + u }\,.
\eeq
The smaller solution, as was found in Section~\ref{subsnumstrong}, determines the location of the 
Strong-Coulomb phase boundary. 
Therefore, single equation \eqref{higgs_phbnd} defines the whole phase diagram
of the theory, see Fig.~\ref{fig:higgsborder}. 
When the deformation parameter $ u $ vanishes, the whole Coulomb/confining phase shrinks to a point
$ m = \Lambda $.

We can now use Eq.~\eqref{higgs_vaceq} to derive a vacuum equation
for the coupling constant $ 2\bren ~=~ |n|^2 $.
From Eq.~\eqref{higgs_numn} one can see that the coupling constant vanishes
at the point \eqref{higgs_cs}, confirming once again that a phase transition takes place.
Plugging Eq.~\eqref{higgs_numn} into Eq.~\eqref{higgs_vaceq}, one arrives at
\beq
\label{higgs_beta}
	\frac{8\pi}{N}\bren ~-~ \ln \left( \frac{8\pi}{N}\bren \,+\, 1\right) 
		~~=~~ 
	\ln\,
	\frac{\mu^2}
        {1 \,+\, u \,+\, \ln \mu^2}\,.
\eeq
At large $u$, one recovers Eq.~\eqref{Hphbetaeq}, and therefore the same analysis of the solutions
sketched after Eq.~\eqref{Hphbetaeq} applies to Eq.~\eqref{higgs_beta} --- there are two solutions,
only one of which is physical.
This is the solution shown in Fig.~\ref{fig:numsignm}.

We now turn to the question of the energy density in the Higgs phase.
In terms of the dimensionless variables, expression \eqref{Veff} can be brought into the form
\begin{align}
\notag
	\frac{4\pi}{N}\,\frac{E_\text{vac}^\text{Higgs}}{\Lambda^2} & ~=~ 
			\frac{1}N\, \sum \bigl| \cs \,-\, \mu^k \bigr|^2\, \ln \bigl| \cs \,-\, \mu^k \bigr|^2 
	\\[1mm]
	&
	~-~ \frac{1}N\, \sum \cs\, \Bigl\{\,  (\mu \,-\, \mu^k) \,+\, (\mu \,-\, \ov{\mu}{}^k) \,\Bigr\}\,
		   		   \ln \lgr 2\mu\cs\, ( 1\, -\, \cos \alpha_k ) \rgr \\[2mm]
\notag
	&
	~-~ ( \cs \,-\, \mu )^2 ~+~ u\, \cs^2\,.
\end{align}
Evaluating the sums for large $N$, and using vacuum equation \eqref{higgs_vaceq}, one comes to the expression
\beq
\label{higgs_energy}
	\frac{4\pi}{N}\,\frac{E_\text{vac}^\text{Higgs}}{\Lambda^2} ~~=~~
	( \mu^2 \,-\, \cs^2 )\, \ln \mu^2 ~-~ (\mu \,-\, \cs)^2 ~-~ u\, \cs^2\,.
\eeq

It is now straightforward to see that this energy density matches the Coulomb phase energy density at the
phase transition point $\mu_*$. 
One eliminates $\cs$ from Eq.~\eqref{higgs_energy} using relation \eqref{higgs_cs}, and resolves the
logarithm $ \ln \mu_*^2 $ via \eqref{higgs_phbnd}.
The result is
\beq
	\frac{4\pi}{N}\,\frac{E_\text{vac}^\text{Coulomb}(\mu_*)}{\Lambda^2} ~~=~~
	\frac{4\pi}{N}\,\frac{E_\text{vac}^\text{Higgs}(\mu_*)}{\Lambda^2} ~~=~~
	\mu_*^4 ~-~ ( 2 \,+\, u ) \mu_*^2 ~+~ 1\,.
\eeq
\begin{figure}
\epsfxsize=11cm
\centerline{\epsfbox{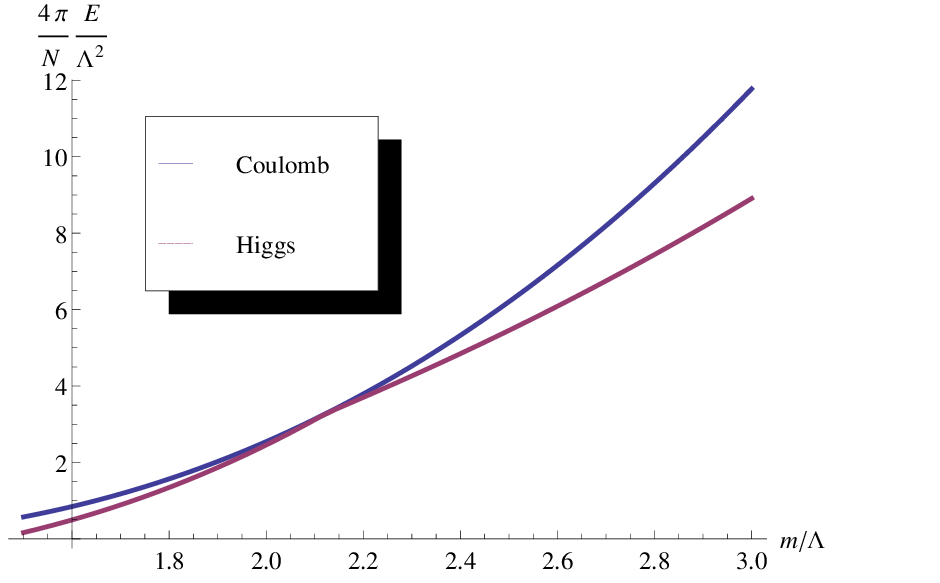}}
\caption{\small 
Energy densities for Coulomb/confining and Higgs phases for $ u = 2 $.
Conjunction occurs exactly at $ \mu = \mu_* $.
}
\label{fig:higgsenergy}
\end{figure}
Figure~\ref{fig:higgsenergy} illustrates this.

\section{More on the Coulomb/confining phase}
\label{moccp}
\setcounter{equation}{0}

As was shown above (Sects.~\ref{scpwbz} and \ref{subshiggph}), both in the strong coupling  and Higgs phases  
the $Z_{2N}$ symmetry is spontaneously broken down to $Z_2$ while in the Coulomb/confining phase this
symmetry remains unbroken (see Sect.~\ref{subscoulco}). In the former two phases we have $N$ degenerate vacua,
while in the later phase the theory has a single vacuum. Just like in non-supersymmetric
\cpn model \cite{GSY05} the vacua   split,  and $N-1$ would-be
vacua become quasivacua, see \cite{SYrev}
for a review. The vacuum splitting can be understood as a manifestation of the Coulomb/confining linear
potential between the kinks \cite{Coleman,W79} that interpolate between the true vacuum, and say,
the lowest quasivacuum. The force is attractive in the kink-antikink pairs leading to  formation
of weakly coupled bound states (weak coupling is the
manifestation of the $1/N$ suppression of the confining potential, see below).
The charged kinks (i.e. the $n$ quanta) are eliminated from the spectrum. This is the reason
why the
$n$ fields were called ``quarks" by Witten \cite{W79}. The spectrum
of the theory consists of $\bar{n} n$-``mesons.'' The picture of
confinement of $n$'s is shown in Fig.~\ref{fig:conf}.

\begin{figure}
\epsfxsize=8cm
\centerline{\epsfbox{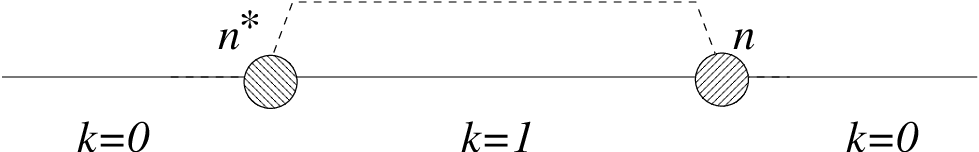}}
\caption{\small 
Linear confinement of the kink-antikink pair.
The solid straight line represents the ground state.
The dashed line shows
the vacuum energy density of the lowest quasivacuum.}
\label{fig:conf}
\end{figure}

This Coulomb/confining linear potential can appear only provided that the photon remains massless,
which is certainly true in the pure bosonic \cpn model. As was pointed out by Witten \cite{W79},
in the supersymmetric \ntwot \cpn model the photon  acquires a nonvanishing
mass due to a chiral coupling to fermions.
In particular, with vanishing twisted masses the photon is massive in both \ntwot \cite{W79} and
\ntwoo \cite{SYhet}. Below we will calculate the photon mass in the model (\ref{bee31}) and
show that it does vanish in the Coulomb/confining phase considered in Sect.~\ref{subscoulco}.
This is in accord with the unbroken $Z_N$ symmetry detected in this phase. It guarantees self-consistency
of the picture.

To this end we start from the
one-loop effective action which is  a function of fields from the gauge supermultiplet
($A_k$, $\sigma$ and $\lambda$). 
After integration over $n^{i}$ and $\xi^i$ in the strong coupling or Coulomb/confining phases
(at $n=0$)
the bosonic part of this effective action takes the form
\cite{SYhet}
\beq
S_{\rm eff}=
 \int d^2 x \left\{
\frac1{4e_{\gamma}^2}F^2_{\mu\nu} + \frac1{e_{\sigma}^2}
|\pt_{\mu}\sigma|^2
+ V(\sigma)+\sqrt{2}(\bar{b}\delta\sigma- b\delta\bar{\sigma})\,F^{*}
  \right\},
\label{effaction}
\eeq
where 
$F^{*}$ is the  dual gauge field strength,
\beq
F^{*}=\frac12\varepsilon_{\mu\nu}F_{\mu\nu}\,,
\eeq
while $V(\sigma)$ can be obtained from (\ref{Veff}) by eliminating $D$ by virtue of its equation of motion (\ref{eff1}). This was done in the closed form for $m=0$ in \cite{SYhet}, see also Sec.~\ref{defefl}.
Here $e^2_{\gamma}$ and  $e^2_{\sigma}$  and $b$ are the coupling constants which
determine the wave function renormalization for  the photon,  $\sigma$ and sigma-photon mixing
respectively ($\delta \sigma$ is the quantum fluctuation of the field $\sigma$ around its VEV). 
These couplings are given by one-loop graphs which we will consider below. In the $m=0$ case
these graphs were calculated in \cite{SYhet}.

The wave-function renormalizations of  the fields from the gauge supermultiplet are, 
in principle, momentum-dependent. We calculate them below in the low-energy limit assuming 
the external momenta to be small.
\begin{figure}
\epsfxsize=6cm
\centerline{\epsfbox{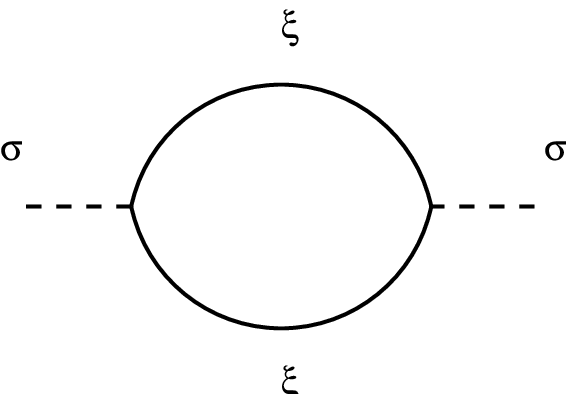}}
\caption{\small
The wave function renormalization for  $\sigma$. }
\label{fig:esigma}
\end{figure}
The wave-function renormalization for $\sigma$ is given by the graph in 
Fig.~\ref{fig:esigma}. A straightforward calculation yields
\beq
\frac1{e^2_{\sigma}}=\frac{1}{4\pi}\,\sum_{i=0}^{N-1}\frac{1}{|\sqrt{2}\sigma+m_i|^2}\,.
\label{esigma}
\eeq
The above graph is given by the integral over
the momenta of the $\xi$ fermions propagating in the loop.
The integral is saturated at momenta of the order of the $\xi$
mass $|\sqrt{2}\sigma+m_i|$.

The wave function renormalization for the gauge field was calculated by Witten in 
\cite{W79} for zero masses. The generalization to the case of nonzero masses takes the form
\beq
\frac1{e^2_{\gamma}}=\frac{1}{4\pi}\,\sum_{i=0}^{N-1}\left[\frac13\,\frac{1}{iD+|\sqrt{2}\sigma+m_i|^2}+
\frac23\,\frac{1}{|\sqrt{2}\sigma+m_i|^2}\right].
\label{egamma}
\eeq
The right-hand side in Eq.~(\ref{egamma}) is given by two graphs in 
Fig.~\ref{fig:photon}, with bosons $n^i$ and fermions $\xi^i$
in the loops. The first term in (\ref{egamma}) comes from bosons while the second
one is due to fermions.
\begin{figure}
\epsfxsize=10cm
\centerline{\epsfbox{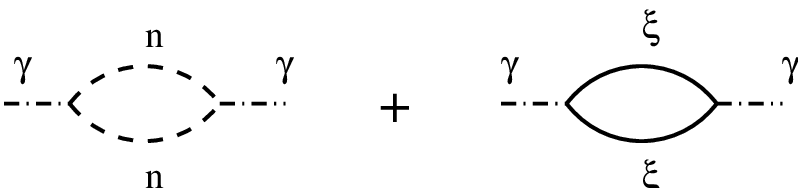}}
\caption{\small
The wave function renormalization for the gauge field.}
\label{fig:photon}
\end{figure}

We see that although both $A_{\mu}$ and $\sigma$ were introduced in (\ref{bee31})
as auxiliary fields (in the limit $e^2_0\to\infty$) after renormalization both 
couplings, $e^2_{\gamma}$ and  $e^2_{\sigma}$, become finite. This makes these
fields physical \cite{W79}.

The $(\delta \sigma)\, F^{*}$ mixing was calculated by Witten in \cite{W79}
in the massless \ntwot theory. This mixing  is due to
the chiral fermion couplings which can make the photon massive in two dimensions. In the effective action
this term is represented by the mixing of the gauge field with the fluctuation of 
$\sigma$. It is given by the graph in Fig.~\ref{fig:photsig}. Direct calculation gives
for the coupling $b$ 
\beq
b= \frac{1}{2\pi}\,\sum_{i=0}^{N-1}\,\frac{\sqrt{2}\sigma+m_i}{|\sqrt{2}\sigma+m_i|^2}\,.
\label{gamma}
\eeq
\begin{figure}
\epsfxsize=6cm
\centerline{\epsfbox{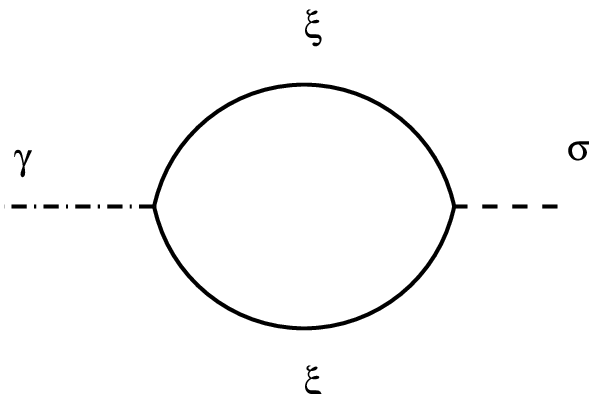}}
\caption{\small
Photon--$\sigma$ mixing. }
\label{fig:photsig}
\end{figure}
With the given value of $b$ the photon mass is
\beq
m_{\rm ph}=  e_{\sigma}e_{\gamma}|b| \,.
\label{phmass}
\eeq

We see that in the strong coupling phase, with   $\sigma\neq 0$, the photon mass does not vanish.
However, 
 in the Coulomb/confining phase in which the VEV of $\sigma$ vanishes,   $b =0$, 
 and 
the photon is massless, as expected,
\beq
m_{\rm ph}=0\, .
\eeq

The above circumstance can be readily understood from symmetry considerations too.
The field $\delta \sigma$ transforms
nontrivially under $Z_{2N}$ symmetry; therefore, the  nonvanishing value of $b$ in (\ref{effaction})
breaks this symmetry. However, $Z_{2N}$ symmetry is restored in the Coulomb/confining phase.
Hence, $b$ should be zero in this phase, and it is.

If the photon mass vanishes,
generating a linear potential, it is not difficult to find the splittings between 
the quasivacua; they are determined by the value of
$e_{\gamma}^2$. From (\ref{egamma}) we  estimate
\beq
E_{k+1}-E_{k}\sim e^2_{\gamma}
=\frac{4\pi}{N}\left\{
\begin{array}{cc}
3\Lambda^2,\;\;\;\;\; m\ll \Lambda\\[2mm]
\frac32 m^2,\;\;\;\;\;\; m\gg \Lambda
\end{array}\,.
\right.
\label{split}
\eeq
We see that the splitting is a $1/N$ effect. This was expected.

\section{Related issues}
\label{relais}
\setcounter{equation}{0}

In this section we  address a few questions which are not necessarily confined to the large-$N$ limit.
Rather, we focus on some general features of our results, with the intention to provide
some useful clarifications/illustrations.

\subsection[Remarks on the mirror representation
for the heterotic CP(1) in the limit of small deformation]
{Remarks on the mirror representation
for the heterotic CP(1) in the limit of small deformation\footnote{Subscript 
$\ssm$ of $\gamma_\ssm$ will indicate that we are working in Minkowski space in this section.}}

In this section we will set all twisted masses to zero.
The geometric representation of  the heterotic \nzt CP(1) model is as follows \cite{SY1}:
\beqn
\label{AAone}
&&
L_{{\rm heterotic}}= 
\zeta_R^\dagger \, i\partial_L \, \zeta_R  + 
\left[\gamma_\ssm \, \zeta_R  \,R\,  \big( i\,\partial_{L}\phi^{\dagger} \big)\psi_R
+{\rm H.c.}\right] \\[4mm]
&&
-g_0^2|\gamma_\ssm |^2 \left(\zeta_R^\dagger\, \zeta_R
\right)\left(R\,  \psi_L^\dagger\psi_L\right)
\nonumber
+
G\, \left\{\rule{0mm}{5mm}
\partial_\mu \phi^\dagger\, \partial^\mu\phi  
+\frac{i}{2}\big(\psi_L^\dagger\!\stackrel{\leftrightarrow}{\partial_R}\!\psi_L 
+ \psi_R^\dagger\!\stackrel{\leftrightarrow}{\partial_L}\!\psi_R
\big)\right.
\nonumber
\\[4mm] 
&&
-\frac{i}{\chi}\,  \big[\psi_L^\dagger \psi_L
\big(\phi^\dagger \!\stackrel{\leftrightarrow}{\partial_R}\!\phi
\big)+ \psi_R^\dagger\, \psi_R
\big(\phi^\dagger\!\stackrel{\leftrightarrow}{\partial_L}\!\phi
\big)
\big]
-
\frac{2(1- g_0^2 |\gamma_\ssm |^2)}{\chi^2}\,\psi_L^\dagger\,\psi_L \,\psi_R^\dagger\,\psi_R
\Big\}\,,
\nonumber
\eeqn
where the field $\zeta_R$ appearing in the first line is
the spinor field on $C$, a  necessary ingredient  of the ${\mathcal N}=(0,2)$ deformation
\cite{EdTo}.
Here $G$ is the metric, $R$ is the Ricci tensor and $\chi \equiv 1+\phi\,\phi^\dagger\,,$
\beq
G=
\frac{2}{g_{0}^2\,\chi^{2}}\,,\qquad R =\frac{2}{\chi^2}\,,
\label{fsmetrone}
\eeq
cf. Eq.~(\ref{eq:RG}).

We assume the deformation parameter $\gamma$ to be small (it is dimensionless)
and work to the leading order in $\gamma$, neglecting $O(\gamma^2)$ effects in the superpotential.
The kinetic terms of the CP(1) fields $\phi$ and $\psi$ contain $\frac{1}{g^2}$ in the normalization
while $\gamma$ in the first line is defined in conjunction with the Ricci tensor, so that there is no $\frac{1}{g^2}$
in front of this term. This convention is important for what follows.

Now, let us remember that the undeformed \ntt  CP(1) model has a mirror representation \cite{MR1,MR2},
a Wess--Zumino model with the superpotential
\beq
{\mathcal W}_{\rm mirror} = \Lambda \left( Y + \frac{1}{Y}\right)\,,
\label{AABone}
\eeq
where $\Lambda$ is the dynamical scale of the CP(1) model.
The question is: ``what is the mirror representation of the deformed model (\ref{AAone}), to the leading order in $\gamma$?"

Surprisingly, this question has a very simple answer.
To find the answer let us observe that the term of the first order in $\gamma$ in (\ref{AAone})
is nothing but the superconformal anomaly in the unperturbed \ntt  model (it is sufficient to consider this anomaly in the unperturbed model since we are after the leading term in $\gamma$ in the mirror representation).
More exactly, in the
\ntt  CP(1) model \cite{ls,ls1}
\beq
\gamma_\mu J^\mu_{\,\,\alpha} = -\frac{ \sqrt 2}{2\pi}\,R\, \left(\partial_\nu \phi^\dagger\right)\left(
\gamma^\nu\psi\right)_\alpha\,,
\label{six}
\eeq
where $J^\mu_{\,\,\alpha}$ is the supercurrent.
In what follows, for simplicity,  numerical factors like $2$ or $\pi$ will be omitted. Equation (\ref{six})
implies that the $O(\gamma )$ deformation term in (\ref{AAone}) can be written as
\beq
\Delta {\mathcal L} = \gamma_\ssm \zeta_R\left(\gamma_\mu \, J^\mu\right)_L
\label{seven}
\eeq 

Since (\ref{six}) has a geometric meaning we can readily rewrite this term in the mirror representation in terms of 
${\mathcal W}_{\rm mirror}$.
 Indeed, in the generalized \nzt Wess--Zumino model the term proportional to $\gamma_\ssm \zeta_R$ is
 \cite{SYneww}
 \beq
 \Delta\cell = \zeta_R \psi_L\,{\mathcal H}^{\,\prime}\,,\qquad {\mathcal H}^{\,\prime} =\partial {\mathcal H}/\partial Y\,,
 \label{arione}
 \eeq
 where ${\mathcal H}$ is the $h$-superpotential.\footnote{It is worth noting
 that in \cite{SYneww} the $h$ superpotential ${\mathcal H}$ was denoted by $S$.
 Note that a broad class of
the $(0,2)$ Landau--Ginzburg models were analyzed, from various perspectives, in \cite{D1,D2,D3}.
The prime interest of these studies was the flow of the  (0,2) Landau-Ginzburg models  to
non-trivial (0,2) superconformal field theories \cite{D1,D2}, and  \nzt analogs of the topological rings
in the \ntt theories \cite{D3}.} Moreover,
 \beq
 \left(\gamma_\mu J^\mu\right)_L = \left({\mathcal W}^{\,\prime} \psi_{L}\right)_{\rm mirror} +O(\gamma)\,.
  \label{aritwo}
 \eeq
Substituting Eq.~(\ref{aritwo}) in (\ref{seven}) and comparing with (\ref{arione}) we conclude that
\beq
{\mathcal H} = \gamma_\ssm\,{\mathcal W}_{\rm mirror}\,.
  \label{arithree}
 \eeq
 In principle, one could have added a constant on the right-hand side, but this would ruin the $Z_2$ 
 symmetry inherent to the \nzt CP(1) Lagrangian. The constant must be set at zero. 
 The scalar potential of the \nzt mirror Wess--Zumino model 
  is \cite{SYneww}
 \beq
 V= |\cw^{\,\prime} |^2+|\mathcal H|^2 = |\cw^{\,\prime}_{\rm mirror} |^2 +|\gamma_\ssm |^2\,|\cw_{\rm mirror} |^2\,.
   \label{arifour}
 \eeq
 where $\cw_{\rm mirror} $ is given in (\ref{AABone}).
 The second equality here is valid in the small-deformation limit.
 
At $\gamma \neq 0$ it is obvious that $V>0$ and supersymmetry is broken.
The $Z_2$ symmetry apparent in (\ref{arifour}) is spontaneously broken too:
we have two degenerate vacua.



\subsection{Different effective Lagrangians}
\label{defefl}

In this section we will comment on the relation between the
effective Lagrangian derived in Sect.~\ref{hecpnsm} from the large-$N$ expansion
 and the Veneziano--Yankielowicz
effective Lagrangian based on  anomalies and supersymmetry.
For simplicity we will set $m_i =0$ in this section. 
Generalization to  $m_i \neq 0$ is straightforward. We  assume the heterotic deformation to be small, $u\ll1$.

The $1/N$ expansion allows one to derive an honest-to-god effective
Lagrangian for the field $\sigma$, valid both in its kinetic and potential parts. The  leading order in
$1/N$ in the potential part is determined by the diagram depicted in Fig.~\ref{fig:esigma}
which gives at zero twisted masses
\beq
{\mathcal L}_{\rm kin} =\frac{N}{4\pi}\,\frac{1}{2|\sigma|^2}\,| \,  \partial_\mu\sigma \,|^2
\,,
\label{fdop1}
\eeq
see (\ref{esigma}).
The  virtual $\xi$ momenta saturating the loop integral are of the order of the $\xi$
mass $\sqrt{2}|\sigma|$. Up to a numerical coefficient this result is obvious since the field
$\sigma$ has mass-dimension 1.

The potential part following from calculations in Sect.~\ref{hecpnsm}
is 
\beq
{\mathcal L}_{\rm pot} = \frac{N}{4\pi}\left\{\Lambda^2 +2|\sigma|^2\left[
\ln\frac{2|\sigma|^2}{\Lambda^2}-1+u\right]
\right\}.
\label{fdop2}
\eeq
All corrections to (\ref{fdop1}) and  (\ref{fdop2}) are suppressed by powers
of $1/N$. For what follows it is convenient to recall the dimensionless variable $ \cs $ (see Eq.~\eqref{csdef}),
\beq
\cs = \frac{ \sqrt{2}\sigma}{\Lambda}\,.
\eeq
Then the large-$N$ effective Lagrangian of the $\sigma$ field takes the form
\beq
{\mathcal L}_{\rm eff} =\frac{N}{4\pi}
\left\{
\frac{1}{\,2\, |\cs|^2}\,\,
 | \partial_\mu\cs \, |^2
+\Lambda^2
\left[1 +|\cs |^2\left( \ln |\cs|^2 -1+u\right)\right]
\right\}.
\label{fdop3}
\eeq

On the other hand, 
the Veneziano--Yankielowicz method \cite{VYan} produces 
an effective Lagrangian in the Pickwick sense. It realizes, in a superpotential,  the  anomalous Ward identities of the 
underlying theory
and other symmetries, such as supersymmetry, and gives no information on the kinetic
part. We hasten to add, though, that in two dimensions
in the undeformed\,\footnote{The key word here is ``undeformed'', i.e. with no heterotic deformation.} CP$(N-1)$ models
the Veneziano--Yankielowicz superpotential ${\mathcal W}_{\rm VY} = \Sigma \ln \Sigma $ (for twisted superfields)
 obtained in \cite{AdDVecSal,ChVa,W93} happens to be exact. This was mentioned above more than once.
  
In terms of the scalar potential for the $\sigma$ field
the Veneziano--Yankielowicz  construction has the form
\beq
V_{VY} = \frac{e^2_\sigma}{2} \left|\frac{N}{2\pi}
\ln\frac{\sqrt{2}\, \sigma}{\Lambda}\right|^2
+\frac{N}{4\pi} u\,2|\sigma|^2\,.
\label{fdop4}
\eeq
The kinetic term (that's where $e^2_\sigma$ comes from) was not determined; however, we can take
it in the form obtained in the large-$N$ expansion, see (\ref{fdop1}),
since it is scale invariant and, hence,  does not violate Ward identities.

Combining 
\beq
e^2_\sigma =\frac{4\pi}{N}\,2|\sigma|^2
\label{fdop5} 
\eeq
(see \cite{SYhet}) with (\ref{fdop4}) we arrive at
\beqn
{\mathcal L}_{VY} 
&=&\frac{N}{4\pi}\,\frac{1}{2|\sigma|^2}\,| \,  \partial_\mu\sigma \,|^2+
 \frac{N}{4\pi}
\left\{ 2\cdot 2|\sigma|^2
 \left|
\ln\frac{\sqrt{2}\, \sigma}{\Lambda}\right|^2
+2|\sigma|^2\, u\right\}
\nonumber\\[3mm]
&=& \frac{N}{4\pi}
\left\{
\frac{1}{\,2\, |\cs|^2}\,\,
 | \partial_\mu\cs \, |^2
 +\Lambda^2
 \left[2|\cs |^2\,\left|\ln \cs\right|^2 + |\cs|^2\,u
 \right]
 \right\}.
 \label{fdop6}
\eeqn
It is obvious that the potential in (\ref{fdop3})
is drastically different from that in (\ref{fdop6}). 
For instance, (\ref{fdop3}) contains a single log, while (\ref{fdop6}) has the square of this logarithm.
We will comment on the difference and the reasons for its appearance 
later. Now, let us have a closer look at the minima of
(\ref{fdop3}) and (\ref{fdop6}). The variable $\cs$ is complex, and there are $N$ solutions
which differ by the phase,
\beq
\cs_* = \left|\cs_* \right|\exp\left(\frac{2\pi k}{N}\right)\,,\qquad k = 0,1, ..., N-1\,,
\eeq
for a more detailed discussion see Ref.~\cite{Kos}.
Each of these solutions represents one of the $N$ equivalent vacua. 
This feature is well-known, and we will omit the phase by setting $k=0$.
Thus, we focus on a real solution.
The minimum of (\ref{fdop3}) lies at
 \beq
 \cs_* = e^{-u/2}
 \label{wdop1}
 \eeq
while the corresponding value of $V_{\rm eff}$ is
\beq
V_{\rm eff} (\cs_* ) = \frac{N}{4\pi}\,\Lambda^2 \left(1- e^{-u}
\right)  .
 \label{wdop2}
\eeq
At the same time, the minimum of (\ref{fdop6}) lies at
\beq
\cs_* = \exp\left(-\frac{1}{2} + \sqrt{\frac{1}{4}-\frac{u}{2}}\right) =  e^{-u/2}\left(1-\frac{u^2}{4} + ...
\right)
\label{wdop3}
\eeq
implying that 
\beqn
V_{\rm VY} (\cs_* )
&=&  \frac{N}{4\pi}\,\Lambda^2 \,\left({1} - \sqrt{{1}-2 {u}} \right)\exp\left(-{1} + \sqrt{{1}-2 {u}}\right)
\nonumber\\[3mm]
& =&
  \frac{N}{4\pi}\,\Lambda^2 \left(1- e^{-u}
\right) \left(1-\frac{u^2}{6}  + ...
\right) .
\label{wdop5}
\eeqn 
The $\sigma$ masses are
\beq
m_\sigma^2 = \left\{
\begin{array}{l}
4\Lambda^2\,e^{-u}\,(1-u)\,,\\[2mm]
4\Lambda^2\,e^{-u}\,(1-u)\,(1-u^2+...)\,,
\end{array}
\right.
\label{wdop4}
\eeq
for (\ref{fdop3}) and (\ref{fdop6}) respectively.
The positions of the minima, the $\sigma$ masses
as well as the vacuum energy densities in these two cases
differ by $O(u^2)$ in relative units. They coincide in the leading and next-to-leading orders in $u$, however.

There are two questions to be discussed: (i) why the effective Lagrangians  (\ref{fdop3}) and (\ref{fdop6}),
 being essentially different, predict identical vacuum parameters in the leading and next-to-leading order in $u$;
 and (ii) why the  parameters extracted from the $1/N$ and Veneziano--Yankielowicz Lagrangians diverge 
 from each other at $O(u^2)$ and  higher orders.

The answer to the first question can be found in \cite{AdDVecSal}. While  the $1/N$ 
Lagrangian is defined unambiguously, the Veneziano--Yankielowicz method determines only the
superpotential part of the action. The kinetic part remains ambiguous. We got used to the fact that
variations of the kinetic part affect only terms with derivatives, which are totally irrelevant for the
potential part. This is not the case in supersymmetry.
The correct statement is that variations of the kinetic part term, in addition to derivative terms,
contains terms with $F\bar F$, which vanish in the vacuum ($F=0$) but alter the form of the potential outside
the vacuum points (minima of the potential). The only requirement to the kinetic term is that
it should obey all Ward identities (including anomalous) of the underlying microscopic theory.
For instance, in the case at hand, the simplest choice $\ln\bar\Sigma\, \ln \Sigma $
does the job. However, 
$$
\ln\bar\Sigma\, \ln \Sigma \left[1+ \frac{(\bar{D}^2 \ln\bar\Sigma)\, ({D}^2\ln \Sigma )}{\bar\Sigma\Sigma }
\right]
$$
does the job as well. In this latter case there is an additional factor
$$
\left[1 +\bar FF/(\bar\sigma^2\sigma^2)   +...  \right] $$
which reduces to 1 in the points where $F=0$ and changes the expression for $F$ (and, hence, the scalar potential)
outside minima (i.e. at $F\neq 0$).

The answer to the second question is even more evident.
The Veneziano--Yan\-ki\-e\-lo\-wicz Lagrangian (\ref{fdop6}) reflects the Ward identities
of the unperturbed CP$(N-1)$ model. That's the reason why the predictions following from this
Lagrangian fail at the level $O(u^2)$, but are valid at the level $O(u)$. 
We remind the reader that it was shown in \cite{SY1} that the vacuum energy density
at the level $O(u)$ is determined by the bifermion condensate in the 
conventional (unperturbed) CP$(N-1)$ model.

One last remark is in order here.
The kinetic term (\ref{fdop1}) is not canonic and singular at $\sigma=0$, implying that this point should be analyzed separately.
One can readily cast (\ref{fdop1}) in the canonic form by a change of variables. Upon this transformation 
$\sigma\to \tilde 
\sigma =2 \ln\sqrt{2}\sigma/\Lambda$ (assuming for simplicity $\sigma$ to be real and positive),
the transformed potential (\ref{fdop2})    develops an extremum at 
$\sigma = 0$ (i.e. $\tilde\sigma \to -\infty$). This extremum is maximum rather than minimum.
Indeed, at $u=0$
\beq
\tilde{\mathcal L}_{\rm pot} = \frac{N\Lambda^2}{4\pi^2}\,(\tilde\sigma - 1)\,e^{\tilde\sigma}
+{\rm const}.
\label{mdop7}
\eeq
It is curious to note that (\ref{mdop7}) exactly coincides with the (two-dimensional)
dilaton effective Lagrangian derived in \cite{SMMS} on the basis of the most general
(anomalous) scale Ward identities.

\subsection{When the \boldmath{$n$} fields can be considered as solitons}
\label{wtnfcb}

Long ago Witten showed \cite{W79} that the $n$ fields in fact describe
kinks interpolating between two neighboring vacua picked up from the set of $N$ degenerate 
supersymmetric vacua
of the \ntt sigma model. The above statement refers to the model with no twisted masses.
(See Fig.~\ref{nkin}).
Here we will discuss the physical status of these states, and the BPS spectrum at large,
as the twisted mass parameter evolves towards large values, $|m|/\Lambda \gg1$.
$N$ will be assumed to be large so that we can use the large-$N$ solutions.

In the undeformed \ntt theory two distinct regimes are known to exist.
\begin{figure}
\epsfxsize=5cm
\centerline{\resizebox{5cm}{!}{\input{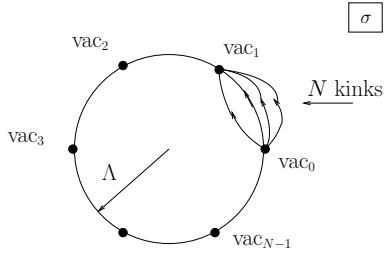}}}
\caption{\small The kinks are represented by the $n$ fields at $|m_i|<\Lambda$. }
\label{nkin}
\end{figure}
At large $m$
the theory is in the Higgs regime, while at small $m$ it is in the
strong coupling regime.
There is no phase transition between the two regimes, since 
the global $Z_N$ symmetry of the model is spontaneously broken in both.
An apparent discontinuity of, say, the derivative of the vacuum expectation of $\sigma$ at
$m=\Lambda$ (see Fig.~\ref{fig22nsigma}) is an artifact of the $N\to\infty$ limit.
However, the BPS spectrum experiences a drastic change in passing from small to large $m$.
In particular, 
at $|m|<\Lambda$, the masses $M_i$ of the $n^i$ fields in the vacuum $|0\rangle$
are
\beq
M_i^2 ~=~ \left| \Lambda \,-\, m_i \right|^2\,,\qquad i\,=\,0,1, ..., N-1\,.
\label{nkma}
\eeq
This follows from Eq.~(\ref{bee31}). The mass degeneracy of the kink $N$-plet is gone
since the twisted mass terms (\ref{two}) break the SU$(N)$ symmetry of the model leaving intact only
U(1)$^{N-1}$ and \zn-symmetries.

On the other hand, at large $m$, in the Higgs regime at weak coupling,  the $n$ fields no longer describe solitons;
rather they represent elementary excitations. In each of the $N$ vacua
there are $2(N-1)$
real elementary excitations. Say, consider the vacuum in which $n^0$ develops a VEV.
The phase of $n^0$ is eaten by the Higgs mechanism providing the mass to the photon field.
The modulus of $n^0$ is excluded from dynamics by the constraint $\left|n^0\right|^2=2\beta$.
The elementary excitations are described by $n^1,\,n^2,\, ... ,\, n^{N-1}$.
The same equation~(\ref{bee31}) implies that the masses of these elementary excitations are\,\footnote{In fact, 
equation $M_k= \left| m_0 - m_k\right|$ is exact.}
(at large $N$),
\beq
M_k^2 = \left| m_0 - m_k\right|^2=
\left(m\, \frac{2\pi k}{N}
\right)^2
\,,\qquad k=1, 2, ..., N-1\,.
\label{nkmap}
\eeq
since the vacuum expectation of $\sigma$ in the vacuum under consideration is $\sigma\approx m_0\equiv m$.
For the values of $k$ that do not scale with $N$, each such mass in (\ref{nkmap}) scales as $1/N$ at large $N$.
This is the consequence of our $Z_N$-symmetric choice of the twisted mass parameters
(\ref{two}). 


What about the kink masses in the Higgs regime? (For brevity we will refer to them as
the weak-coupling regime (WCR) kinks, although this is not quite precise.
Indeed, if $m$ is in the Higgs domain but $\sim {\rm a\,\, few}\times\Lambda$, the coupling constant is 
of order 1.) The masses of the WCR
kinks
can be found from the Veneziano--Yankielowicz
effective superpotential \cite{HaHo},
which is exact in the model under consideration,
\beq
\cw_{\rm VY} 
=
 \frac{1}{4\pi}
\,\sum_{i=0}^{N-1}\left(\sqrt{2}\,\sigma -m_i
\right)\ln\frac{\sqrt{2}\,\sigma -m_i}{\Lambda}-\frac{N}{4\pi}\sqrt{2}\,\sigma\,.
\label{widu}
\eeq
For definiteness $N$ is assumed to be even.
One should be careful with the logarithmic function. It is obvious that (\ref{widu})
is defined up to $im_p/2$ times any integer number.
For the kinks interpolating between the  vacua $|0\rangle$ and  $|p\rangle$  one has \cite{HaHo}
\beq
M_{\rm kink} 
=
 2\left|\Delta \cw\right|= 2\left|\cw_{\rm VY} (\sigma_0) - \cw_{\rm VY} (\sigma_p)+\frac{i}{2}\left(m_0-m_p\right) q  \right|\,,
   \label{widup}
\eeq
where the logarithmic function in (\ref{widu}) is defined with the cut along the negative semi-axis, and $q$ is an arbitrary integer.
Since the vacuum values of sigma satisfy
$$(\sqrt{2}\sigma)^N = m^N + \Lambda^N$$ and $|m/\Lambda| >1$, we 
have $\sqrt{2}\sigma_p = m_p$, with the exponential accuracy.
If  $\sigma_0=m_0=m$ and $|p|<N/4$, in calculating  $\cw_{\rm VY} (\sigma_0, \,\sigma_p)$
we do not touch the cut of the logarithm, and then we can use the following expression\,\footnote{One might be tempted to use Eq.~(\ref{usefulformula}) here, but this does not work
since in this expression we have a different argument of the logarithm.}
\beqn
\cw_{\rm VY}(\sigma)
&=&
  \frac{1}{4\pi}\,\sqrt{2}\sigma\,   \ln\left(\frac{\sqrt{2}\sigma}{e\Lambda}\right)^N
+\frac{1}{4\pi\,N }\,\sqrt{2}\sigma\,F_N(y)\,,
\nonumber\\[3mm]
F_N(y) 
&\equiv&
 \sum_{k=1}^\infty \,\,\frac{N}{k\, (kN-1)}\, y^k
\nonumber\\[3mm]
 y 
&=& 
\left(\frac{m}{\sqrt{2}\,\sigma}\right)^N\,.
\label{chet1}
\eeqn
Note that $F_N$ is finite at $N\to \infty$ and $|y|\leq 1$. 
As a result, at  $|p|<N/4$ we arrive at
\beq
M_{\rm kink} 
=
\frac{N\,m}{\pi}\,\sin\frac{\pi\, |p|}{N}\left|\ln\frac{m}{e\Lambda} +2\pi i\frac{q}{N} 
\right|\,.
\label{chet4}
\eeq
The lightest mass is obtained by setting $q=0$,
\beq
M_{{\rm kink}*} 
=
\frac{N\,m}{\pi}\,\sin\frac{\pi\, |p|}{N}\left|\ln\frac{m}{e\Lambda} 
\right|
\label{chet4w}
\eeq
At $p\sim N^0$ and
$|m/\Lambda |\gg1$ the kink mass does not scale with $N$. In addition,  it is enhanced by 
the large logarithm $\ln m/e\Lambda\sim (Ng^2)^{-1}$ compared to the masses of the elementary 
excitations. If $p\sim N^1$, the scaling law is
$$
M_{\rm kink}\sim N\,m \, \ln(m/\Lambda )\,,
$$
cf. Eq.~(\ref{kinkmass}).
Each kink has a tower of excitations on top of it, corresponding to $q\neq 0$.
The existence of excitations is due to the fact that
in addition to  the topological charge, the kinks can  have U(1) charges \cite{Dor}. 
Excitations decay on the curve of marginal stability (CMS) $|m| =e\Lambda$, see \cite{Olmez}.


\section{Conclusions}
\label{conclu}

In this paper we presented the large-$N$ solution of the
two-dimensional  heterotic $\mbox{\ntwoo}$
\cpn model. Our studies were motivated by the fact
that this model emerges on the world sheet of non-Abelian strings
supported in a class of four-dimensional \none Yang--Mills theories.
The non-trivial dynamics which we observed -- with three distinct phases, confinement and no confinement,
 and two phase transitions --
must somehow reflect  dynamics of appropriate four-dimensional
 theories. If so, we open a window to a multitude of
unexplored dynamical scenarios in \none theories. But this is a topic for a separate investigation.

The heterotic CP models in two dimensions are of great interest on their own. For instance, 
heterotically deformed weighted 
CP models with twisted masses exhibit even a richer phase diagram with highly nontrivial phases.
The study of these models in the large-$N$ limit is in full swing now \cite{last}.

\section{Erratum}

There is a typographical error in Eq. (C8). This equation should coincide with (E27).

Equation (E27) was obtained from normalization of the $\bar\zeta i\partial_L\zeta$ term
in the first line in the first (unnumbered) equation on page 9 of \cite{onea} (normalization to unity). 
The coefficient of this term was not established to order $O(\delta^2)$; the absence of $O(\delta^2)$ was not verified. 

In fact, the requirement of holomorphy discussed in \cite{twoa}, implies that an $O(\delta^2)$ correction is present in the coefficient in front of the $\bar\zeta i\partial_L\zeta$ term on page 9 of \cite{onea}, implying that both Eqs. (E27) and (C8) have no denominators, i.e.
\beq
\tilde\gamma_{(E)} = \sqrt 2\, \delta\,.
\eeq

\section*{Acknowledgments}

We are grateful to A. Gorsky, Andrey Losev, Victor Mikhailov, and A. Vainshtein 
for very useful discussions. We would like to thank our referees for a number of suggestions aimed at improving
presentation of our results.

The work of PAB was supported in part by the NSF Grant No. PHY-0554660. 
PAB is grateful for kind hospitality to FTPI, University of Minnesota, where a part of this work was done. 
The work of MS was supported in part by DOE grant DE-FG02-94ER408. 
The work of AY was  supported 
by  FTPI, University of Minnesota, 
by RFBR Grant No. 09-02-00457a 
and by Russian State Grant for 
Scientific Schools RSGSS-11242003.2.

\newpage

\addcontentsline{toc}{section}{Appendices}
\setcounter{section}{0}
\renewcommand{\thesection}{\Alph{section}}

\section{Notation in Euclidean Space}
\label{app:eucl}
%
%

\setcounter{equation}{0}

Since \cpn sigma model can be obtained as a dimensional reduction from four-dimensional theory,
we present first our four-dimensional notations.
The indices of four-dimensional spinors are raised and lowered by the SU(2) metric tensor,
\beq
	\psi_\alpha ~=~ \epsilon_{\alpha\beta}\, \psi^\beta, \qquad
	\ov{\psi}{}_{\dot{\alpha}} ~=~ \epsilon_{\dot{\alpha}\dot{\beta}}\, \ov{\psi}{}^{\dot\beta}, \qquad 
	\psi^\alpha ~=~ \epsilon^{\alpha\beta}\, \psi_\beta, \qquad
	\ov{\psi}{}^{\dot{\alpha}} ~=~ \epsilon^{\dot\alpha\dot\beta}\, \ov{\psi}{}_{\dot\beta}~,
\eeq
	where
\beq
	\epsilon_{\alpha\beta} ~=~ \epsilon_{\dot\alpha\dot\beta} ~=~
			\lgr \begin{matrix}
			     	\ 0\  &  \ 1\   \\
				 -1\  &  \ 0\  
			     \end{matrix} \rgr,
	\qquad \text{and} \qquad
	\epsilon^{\alpha\beta} ~=~ \epsilon^{\dot\alpha\dot\beta} ~=~
			\lgr \begin{matrix}
				\ 0\ &   -1\   \\
				\ 1\ &  \ 0\ 
			     \end{matrix} \rgr.
\eeq
The contractions of the spinor indices are short-handed as
\beq
	\lambda\psi ~=~ \lambda_\alpha\, \psi^\alpha\,, \qquad
	\ov{\lambda\psi} ~=~  \ov{\lambda}{}^{\dot\alpha}\, \ov{\psi}{}_{\dot\alpha}\,.
\eeq
The sigma matrices for the euclidean space we take as
\beq
	\sigma^{\alpha\dot\alpha}_\mu ~=~  \lgr {\bf 1},\quad -i\,\tau^k \rgr^{\alpha\dot{\alpha}}
	\hspace{-2.0ex},
	\qquad
	\ov{\sigma}{}_{\dot\alpha\alpha\, \mu} ~=~ 
			\lgr {\bf 1},\quad i\, \tau^k \rgr_{\dot\alpha\alpha},
\eeq
where $ \tau^k $ are the Pauli matrices.

Reduction to two dimensions can be conveniently done by picking out $ x^0 $ and $ x^3 $ 
as the world sheet (or ``longitudinal'') coordinates, and integrating over the orthogonal coordinates. 
The two-dimensional derivatives are then defined to be
\beq
	\p_R  ~~=~~ \p_0 ~+~ i\p_3\,, \qquad   \p_L ~~=~~ \p_0 ~-~ i\p_3\,.
\eeq
One then identifies the lower-index spinors as the two-dimensional left- and right-handed chiral spinors
\beq
	\xi_{R} ~=~ \xi_{1}\,, \quad\qquad
	\xi_{L} ~=~ \xi_{2}\,, \quad\qquad\qquad
	\ov{\xi}{}_{R} ~=~ \ov{\xi}{}_{\dot{1}}\,, \quad\qquad
	\ov{\xi}{}_{L} ~=~ \ov{\xi}{}_{\dot{2}}\,.
\eeq
With these assignments, the dimensional reduction for the contracted spinors then takes the following form
\beq
	\xi_\alpha\, \lambda^\alpha   ~=~   -\, \xi_{[R}\, \lambda_{L]}\,,
	\qquad\qquad
	\ov{\xi^{\dot\alpha}\, \lambda}{}_{\dot\alpha}   ~=~  \ov{\xi}{}_{[R}\, \ov{\lambda}{}_{L]}\,.
\eeq

For two-dimensional variables, the CP($N-1$) indices are written as upper ones 
\[
	n^l\,, \quad \xi^l\,,
\]
and as lower ones for the conjugate moduli
\[
	\ov{n}{}_l\,, \quad \ov{\xi}{}_l\,, 
\]
where $ l~=~1,\, ...,\, N $.
In the geometric formulation of CP($N-1$), global indices are written upstairs in both cases, only
for the conjugate variables the indices with bars are used 
\[
	\phi^i\,,\ \psi^i\,, \qquad \ov{\phi}{}^\bi\,,\ \ov{\psi}{}^\bi\,, 
	\qquad\qquad i,\ \bi ~=~ 1,\,...,\,N-1\,,
\]
and the metric $ g_{i\bj} $ is used to contract them.


\section{Minkowski versus Euclidean formulation}
 \renewcommand{\theequation}{\Alph{section}.\arabic{equation}}
\setcounter{equation}{0}
 
 \renewcommand{\thesubsection}{\Alph{section}.\arabic{subsection}}
\setcounter{subsection}{0}
\label{app:mink}

In the bulk of the paper we use both, Minkowski and Euclidean conventions.
It is useful to summarize the transition rules.
If the Minkowski coordinates are
\beq
x^\mu_M =\{t,\,z\}\,,
\label{appeone}
\eeq
the passage to the Euclidean space requires
\beq
t \to - i\tau\,,
\label{appe2}
\eeq
and the Euclidean coordinates are
\beq
x^\mu_M =\{\tau,\,z\}\,.
\label{appe3}
\eeq
The derivatives are defined as follows:
\beqn
\pt_L^M &=& \pt_t+\pt_z\,,\qquad \pt_R^M = \pt_t- \pt_z\,,
\nonumber\\[2mm]
\pt_L^E &=& \pt_\tau - i \pt_z\,,\qquad \pt_R^E = \pt_\tau + i \pt_z\,.
\label{appe4}
\eeqn
The Dirac spinor is
\beq
\Psi =\left(
\begin{array}{c}
\psi_R\\[1mm]
\psi_L
\end{array}
\right)
\label{appe5}
\eeq
In passing to the Euclidean space $\Psi^M = \Psi^E$;
however, $\bar\Psi$ is transformed,
\beq
\bar\Psi^M \to i \bar \Psi^E\,.
\label{appe6}
\eeq
Moreover, $\Psi^E$ and $\bar \Psi^E$ are {\em not} related by the complex conjugation operation.
They become independent variables. The fermion gamma matrices are defined as
\beq
\bar\sigma^\mu_M =\{1,\,-\sigma_3\}\,,\qquad \bar\sigma^\mu_E =\{1,\, i\sigma_3\}\,.
\label{appe7}
\eeq
Finally, 
\beq
\cell_E =- \cell_M (t=-i\tau , ...).
\eeq
With this notation, formally, the fermion kinetic terms in $\cell_E $ and $\cell_M $
coincide.


\section{Parameters of heterotic deformation}
 \renewcommand{\theequation}{\Alph{section}.\arabic{equation}}
\setcounter{equation}{0}
 
 \renewcommand{\thesubsection}{\Alph{section}.\arabic{subsection}}
\setcounter{subsection}{0}
 \renewcommand{\thetable}{\Alph{section}.\arabic{table}}
\label{app:het}

In this section we list the definitions of parameters of heterotic deformations
used in previous papers, and their relations to each other, including relations
between the corresponding parameters in Minkowski and Euclidean spaces.

It is reasonable to first collect the constants and couplings which necessarily
accompany the heterotic deformation in theories in which the \ntwoo CP($N-1$) model arises.
If the CP($N-1$) model is obtained from a four-dimensional bulk theory, then its coupling
will be related to the non-abelian gauge coupling of the latter theory
\beq
	2\beta ~~=~~ \frac{4\pi}{g_2^2}\,.
\eeq
From the point of view of the two-dimensional theory, $ \beta $ can be understood as the 
radius of the CP($N-1$) space
\beq
	2\beta ~~=~~ \frac{2}{g_0^2}\,,
\qquad\qquad
K ~=~ 
\frac{2}{g_{0}^{2}}\ln\left(1+\sum_{i,\bar j=1}^{ N-1}\bar\Phi^{\,\bar j}\delta_{\bar j i}\Phi^{i}\right).
\eeq
The latter K\"ahler potential induces the round Fubini-Study metric
\beq
G_{i\bj} ~=~ \frac{\p^2\, K}{\p\phi^i\, \p\ov{\phi}{}^\bj}\,.
\eeq

Table \ref{app:par} displays the equivalent definitions of the heterotic deformations
(the bifermionic terms). 
The subscripts relate the corresponding terms to Minkowski or Euclidean space.
Equivalence in most cases means equality, except for the relations between the Minkowski and
Euclidean space expressions, for which the Lagrangians are defined differently, see Appendix~\ref{app:mink}.
For reasons explained below, different normalizations of the fields involved
in the bifermionic coupling --- $ n^l $, $ \xi^l $ and $ \zr $ --- 
were used in definitions of different parameters. 
In the table, the fields that are normalized canonically are shown in bold face, otherwise
the fields ({\it i.e.} their kinetic terms) are normalized to $ 2\beta $.

\begin{table}[h]
\begin{center}
\begin{tabular}{cccc}
 parameter & deformation term & $O(N)$ scaling  &  defined in \\[0.3cm]
\hline
\hline
\\[-0.3cm]
$\delta$  & $ 2\beta \cdot 2i\, \delta\, \blal \zr $   & $ O(1) $ &  \cite{EdTo,SY1} \\[0.3cm]
$\gamma_\ssm$ & $ \gamma_\ssm\,g_0^2\, \boldsymbol{\zr}\, G_{i\bj} (i \p_L \bphi^\bj)\psi_R^i $ & 
$ \sqrt{N} $ & \cite{SY1} \\[0.3cm]
$ \gamma_\sse $ & $ \gamma_\sse ~=~ i\, \gamma_\ssm $ & $ \sqrt{N} $ & --- \\[0.3cm] 
$\wt{\gamma}{}_\sse$ & $ 2\beta \cdot \wt{\gamma}{}_\sse\, (i\p_L \nbar) \xir \zr $   & $ O(1) $   & \cite{BSY1} \\[0.3cm]
$\omega$  & $ 2i\, \omega\, \boldsymbol{\blal} \boldsymbol{\zr} $  & $ \sqrt{N} $ & \cite{SYhet} \\[0.3cm]
$u$    & $ \frac{N}{2\pi}\, u\, |\sigma|^2 $  & $O(1)$ & \cite{SYhet} \\[0.3cm]
\end{tabular}
\caption{Parameters of heterotic deformation of the CP($N-1$) theory.}
\label{app:par}
\end{center}
\end{table}
One useful normalization (this is {\it not} the normalization used in this paper) for the 
supertranslational and superorientational moduli is $ 2\beta $,
\beq
  \mc{L} ~~=~~ 2\beta \lgr |\p n|^2 ~+~ \bxir\, i\p_L\, \xir ~+~ \bzr\, i\p_L\, \zr ~+~ \dots \rgr.
\eeq
In this normalization one has $ |n|^2 ~=~ 1 $.
In particular, this naturally comes out when the two-dimensional sigma model is obtained from
a four-dimensional bulk theory.
This normalization is useful for studying the sigma model classically. 
The \ntwoo deformation of the sigma model in the gauge formulation is done via the \ntwoo superpotential
\beq
	\hat{\mc{W}}(\sigma) ~=~ \delta\,\frac{\sigma^2}{2}\,.
\eeq
The latter formula is schematical only, as we do not use \ntwoo superfields in this paper. 
Rather, the superpotential leads to the couplings of the type
\beq
\label{Wterms}
	2\beta \lgr 2i\, \delta\, \blal \zr ~+~ 2i\, \ov{\delta}\, \bzr\, \lal ~+~ \dots \rgr .
\eeq
The deformation parameter $ \wt{\gamma} $ was defined in the $|n|^2 ~=~ 1 $ normalization,
\beq
\label{defwtgamma}
	\mc{L}  ~\supset~  2\beta \lgr |\p n|^2 ~+~ \bzr\, i\p_L\, \zr ~+~ \dots 
				~+~ \wt{\gamma}{}_\sse\, (i\p_L \nbar) \xir \zr \rgr 
\eeq
and is related to the above superpotential deformation $ \delta $ as,
\beq
\label{gammadelta}
	\wt{\gamma}{}_\sse  ~~=~~  \frac{\sqrt{2}\,\delta}{ 1 ~+~ 2 |\delta|^2 }\,.
\eeq
Another form of the bifermionic mixing term is found in CP(1) model, where it can be written in
terms of real variables $ S^a $ and $ \chi^a $, see \cite{SY1},
\beq
\label{hetcp1}
	\mc{L} ~~=~~ \beta \lgr
			\frac{1}{2} (\p_\mu S^a)^2 ~+~ \frac{1}{2}\,\chi_R^a i\p_L \chi_R^a 
				     ~+~ 2\, \bzr i\p_L\, \zr  
				     ~+~ \wt{\gamma}{}_\sse\, \chi_R^a (i\p_L S^a) \zr ~+~ \dots \rgr.
\eeq
The relation between $ S^a $ and $ n^l $ (also $ \chi^a $ and $ \xi^l $) depends on the 
normalization of the latter, and in the $ | n |^2  ~=~ 1 $ case takes the form
\beq
	S^a  ~~=~~ \nbar\, \tau^a n\,,  \qquad\qquad  
	\chi^a_{L,R} ~~=~~ \nbar\,\tau^a \xi_{L,R} ~+~ \bxi_{L,R}\, \tau^a n\,.
\eeq
Note the extra factors of $1/2$ and $2$ required for precise matching 
between Eqs.~\eqref{hetcp1} and \eqref{defwtgamma}.

The $ 2\beta $ normalization of the kinetic terms also arises in the geometric formulation 
of the heterotic CP($N-1$) model.
In the latter case, however, one would argue, that the supertranslational variable $ \zr $ has
nothing to do with geometry, and therefore need not have $ 2\beta $ in front of its kinetic term.
Parameter $\gamma$ was originally introduced in \cite{SY1} in the geometric formulation,
in Minkowski space,
\begin{align}
\notag
	\mc{L} & ~~\supset~~ \bm{\bzr}\, i\p_L\, \bm\zr ~+~
				G_{i\bj}\, \lgr \p_\mu \bphi^\bj\, \p_\mu \phi^i   ~+~
						\bpsi^\bj\, \gamma^\mu D_\mu \psi^i \rgr \\[1mm]
		& ~~~+~ \gamma_\ssm\,g_0^2\, \boldsymbol{\zr}\, G_{i\bj} (i \p_L \bphi^\bj)\psi_R^i 
		~+~ \dots\,.
\end{align}
Here, again, the bold face of the variable $ \bm\zr $ indicates that its kinetic term is 
normalized canonically, whereas the kinetic terms of the orientational variables are naturally 
supplied with a factor of $ 2\beta $ coming from the metric $ G_{i\bj} $.
The correspondence between the Minkowski and Euclidean space Lagrangians gives
\beq
\label{app:gammamink}
	\gamma_\ssm  ~=~  -i\, \gamma_\sse\,,
\eeq
with
\beq
         \gamma_\sse ~=~ \frac{1}{\sqrt{2}g_0} \wt\gamma{}_\sse 
		      ~=~ \sqrt{\frac{\beta}{2}}\, \wt\gamma{}_\sse\,.
\eeq

Gauge formulation appears to be the most useful for studying the quantum effects of the sigma
model.
In the quantum theory, one prefers to depart from $ |n|^2 $ being equal to unity, as the former determines the coupling
of the theory and may, in principle, vanish. 
One absorbs all outstanding factors of $ 2\beta $ into the definition of $ n^l $, $ \xi^l $
and $ \zr $.
{\it This is} the normalization used in the bulk of this paper,
\beq
	\mc{L} ~=~ | \nabla_\mu \bm {n}{}^l |^2   ~+~ \bm{\bxi}{}_l\, i\ov{\sigma}{}^\mu\nabla_\mu\, \bm{\xi}{}^l ~+~ 
		   \bm\bzr\, i\p_L\, \bm\zr ~+~ 
		   \frac{1}{\sqrt{2\beta}}\, \wt{\gamma}{}_\sse\, \, (i\p_L \bm \nbar) \bm\xir \bm\zr 
		   ~+~ \dots\,,
\eeq
where $ \wt{\gamma}{}_\sse $ is still related to $ \delta $ via Eq.~\eqref{gammadelta}.
The \ntwoo superpotential couplings \eqref{Wterms} become
\beq
		 2i\, \omega\, \blal \bm\zr ~+~ 2i\, \ov{\omega}\, \bm\bzr\, \lal ~+~ \dots \,,
\eeq
where
\beq
\label{omegadelta}
		\omega ~~=~~ \sqrt{2\beta}\, \delta\,.
\eeq
In the large-$N$ limit, a more appropriate parameter appears to be $ u $,
\beq
		u ~~=~~ \frac{8\pi}{N}\, |\omega|^2\,,
\eeq
which enters the effective potential as
\beq
		V_\text{eff}  ~~\supset~~  \frac{N}{2\pi} \cdot u\, |\sigma|^2\,.
\eeq

Note, that relation \eqref{omegadelta} is classical, while quantum mechanically, all couplings run.
In the one-loop approximation of this paper, the couplings $ \omega $ and $ \beta $ diagonalize the
renormalization group equations, and are, therefore, the appropriate parameters in the
quantum regime. 
In the large-$N$ limit, again, parameter $ u $ is more preferrable. 

Finally, in Ref.~\cite{SY1}, parameter $ \alpha $ is used, which relates to $ \delta $ and $ \gamma $ via
\beq
	\alpha   ~=~ \frac{\delta}{\sqrt{1 ~+~ |\delta|^2} }\,,
	\qquad\qquad
	\gamma_\sse ~=~ \sqrt{\frac{\beta}{2}}\, \wt{\gamma}{}_\sse  
                    ~=~ \sqrt{\beta}\,\frac{\alpha}{\sqrt{1 \,+\, |\alpha|^2}}\,.
\eeq

The relation between the heterotic parameters in Euclidean and Minkowski spaces 
is given by Eq.~\eqref{app:gammamink}.
Everywhere in the paper where there is no menace of confusion we omit the super/sub\-scripts $M,E$.


\section
[Global symmetries of the CP\boldmath{$(N-1)$} model with \boldmath{$Z_{N}$}-symmetric
 twisted masses]
{Global symmetries of the CP\boldmath{$(N-1)$} model with \boldmath{$Z_{N}$}-symmetric
 twisted masses\,\footnote{See also the Appendix in Ref.~ \cite{Shifman:2009ay}.}}
 \renewcommand{\theequation}{\Alph{section}.\arabic{equation}}
\setcounter{equation}{0}
 \renewcommand{\thesubsection}{\Alph{section}.\arabic{subsection}}
\setcounter{subsection}{0}
%
\label{app:symm}

 In the absence of the twisted masses
the model is SU$(N)$ symmetric. The twisted masses (\ref{two}) explicitly break this symmetry 
of the Lagrangian (\ref{bee31}) down to U$(1)^{N-1}$,
\beqn
n^\ell&\to& e^{i\alpha_\ell}n^\ell\,,\quad \xi^\ell_R \to e^{i\alpha_\ell}\xi^\ell_R\,
\quad \xi^\ell_L \to e^{i\alpha_\ell}\xi^\ell_L\,,\quad \ell=1,2, ..., N\,,
\nonumber\\[2mm]
\sigma
&\to&
 \sigma\,,\quad \lambda_{R,L}\to \lambda_{R,L}\,.
 \label{appe9}
\eeqn
where $\alpha_\ell$ are $N$ constant phases different for different $\ell$. 

Next, there is a global vectorial U(1) symmetry which rotates all fermions $\xi^\ell$
in one and the same way, leaving the boson fields intact,
\beqn
\xi^\ell_R 
&\to& 
e^{i\beta}\xi^\ell_R\,, \quad
 \xi^\ell_L \to e^{i\beta}\xi^\ell_L\,,\quad \ell=1,2, ..., N\,,
\nonumber\\[2mm]
\lambda_R 
&\to&
 e^{-i\beta}\lambda_R\,,\quad 
\lambda_L \to e^{-i\beta}\lambda_L\,,
\nonumber\\[2mm]
n^\ell &\to& n^\ell\,,\quad \sigma\to\sigma\,.
\label{appeten}
\eeqn

Finally, there is a discrete $Z_{2N}$ symmetry which is of most importance for our purposes.
Indeed, let us start from the axial U$(1)_R$ transformation which would be a symmetry
of the classical action at $m=0$ 
 (it is anomalous, though, under quantum corrections),
\beqn
\xi^\ell_R 
&\to& 
e^{i\gamma}\xi^\ell_R\,, \quad
 \xi^\ell_L \to e^{-i\gamma }\xi^\ell_L\,,\quad \ell=1,2, ..., N\,,
 \nonumber\\[2mm]
 \lambda_R 
&\to&
 e^{i\gamma}\lambda_R\,,\quad 
 \lambda_L \to e^{-i\gamma}\lambda_L\,,\quad \sigma \to e^{2i\gamma}\sigma\,,
\nonumber\\[2mm]
n^\ell
&\to&
 n^\ell\,.
 \label{appe11}
\eeqn
With $m$ switched on and the chiral anomaly included, this transformation 
is no longer the symmetry of the model. However, a discrete $Z_{2N}$ subgroup survives both the inclusion of anomaly and $m\neq 0$. This subgroup corresponds to
\beq
\gamma_k =\frac{2\pi i k}{2N}\,,\quad k= 1,2, ..., N\,.
\label{appe12}
\eeq
with the simultaneous shift
\beq
\ell\to \ell - k\,.
\label{appe13}
\eeq
In other words,
\beqn
\xi^\ell_R 
&\to& 
e^{i\gamma_k}\xi^{\ell-k}_R\,, \quad
 \xi^\ell_L \to e^{-i\gamma_k }\xi^{\ell-k}_L\,, 
 \nonumber\\[2mm]
 \lambda_R 
&\to&
 e^{i\gamma_k}\lambda_R\,,\quad 
 \lambda_L \to e^{-i\gamma_k}\lambda_L\,,\quad \sigma \to e^{2i\gamma_k}\sigma\,,
 \nonumber\\[2mm]
 n^\ell &\to & n^{\ell-k}\,.
 \label{bee35}
\eeqn
This $Z_{2N}$ symmetry  relies on the particular choice of masses 
given in (\ref{two}).

When we switch on the heterotic deformation, the \zn\, transformations (\ref{bee35}) must be supplemented by
\beq
\zeta_R \to  e^{-i\gamma_k }\, \zeta_R\,.
\label{bee35p}
\eeq
The symmetry of the Lagrangian (\ref{sigma_phys}) remains intact.

The order parameters for the $Z_N$ symmetry are as follows:
(i) the set of the vacuum expectation values
$\{ \langle n^0\rangle,\,\, \langle n^1\rangle, \,...\, \langle n^{N-1}\rangle\}$
and (i) the bifermion condensate $\langle  \bar\xi_{R,\,\ell}\,\xi^\ell_L\rangle$.
Say, a nonvanishing value of $\langle n^0\rangle$ or  $\langle  \bar\xi_{R,\,\ell}\,\xi^\ell_L\rangle$ implies that the $Z_{2N}$ symmetry of the action is broken down to
$Z_2$. The first order parameter is more convenient for detection
at large $m$ while the second at small $m$. 

It is instructive to illustrate the above conclusions
in  the geometrical  formulation of the sigma model. 
Namely, in components the Lagrangian of the model is
(for simplicity we will consider CP(1); generalization to CP$(N-1)$ is straightforward)
\beqn
&&
{\mathcal L}_{\,  CP(1)}= G\, \Big\{
\partial_\mu \bar{\phi}\, \partial^\mu\phi -|m|^2{\bar{\phi}\,\phi} 
+\frac{i}{2}\big(\psi_L^\dagger\!\stackrel{\leftrightarrow}{\partial_R}\!\psi_L 
+ \psi_R^\dagger\!\stackrel{\leftrightarrow}{\partial_L}\!\psi_R
\big)
\nonumber\\[1mm] 
&&
-i\,\frac{1-\bar{\phi}\,\phi}{\chi} \,\big(m\,\psi_L^\dagger \psi_R + \bar m
\psi_R^\dagger \psi_L
\big)
\nonumber\\[1mm] 
&&
-\frac{i}{\chi}\,  \big[\psi_L^\dagger \psi_L
\big(\bar{\phi} \!\stackrel{\leftrightarrow}{\partial_R}\!\phi
\big)+ \psi_R^\dagger\, \psi_R
\big(\bar{\phi}\!\stackrel{\leftrightarrow}{\partial_L}\!\phi
\big)
\big]
\nonumber\\[1mm]
&&
-
\frac{2}{\chi^2}\,\psi_L^\dagger\,\psi_L \,\psi_R^\dagger\,\psi_R
\Big\}\,,
\label{Aone}
\eeqn
where 
\beq
\chi = 1+\bar{\phi}\,\phi\,,\quad G= \frac{2}{g_0^2\,\chi^2}\,.
\eeq
The $Z_2$ transformation corresponding to (\ref{bee35}) is
\beq
\phi \to -\frac{1}{\bar{\phi}}\,,\qquad \psi_R^\dagger \psi_L\to -
\psi_R^\dagger \psi_L\,.
\label{bee40}
\eeq
 The order parameter which can detect breaking/nonbreaking of the above
symmetry is
\beq
\frac{m}{g_0^2} \left(1- \frac{g_0^2}{2\pi}
\right)\, \frac{\bar{\phi}\,\phi-1}{\bar{\phi}\,\phi+1} - 
i R \psi_R^\dagger \psi_L\,.
\eeq
Under the transformation (\ref{bee40}) this order parameter changes sign.
In fact, this is  the central charge of the \ntwo
sigma model, including the anomaly  \cite{ls,ls1}.

\section{Geometric formulation of the model}
 \renewcommand{\theequation}{\Alph{section}.\arabic{equation}}
\setcounter{equation}{0}
 \renewcommand{\thesubsection}{\Alph{section}.\arabic{subsection}}
\setcounter{subsection}{0}
\label{app:geom}

Here we will  briefly review the \ntt supersymmetric \cpn models in the
geometric formulation. 

\subsection{Geometric formulation, \boldmath{$\tilde\gamma=0$} }

As usual, we start from the undeformed case.
The target space is the $N-1$-dimensional K\"ahler manifold 
parametrized by the fields $\phi^{i},\,\phi^{\dagger\,\bar j}$, $\,i,\bar j=1,\ldots,N-1$,
which are the lowest components of the chiral and antichiral superfields 
\beq
\Phi^{i}(x^{\mu}+i\bar \theta \gamma^{\mu} \theta),\qquad \bar\Phi^{\bar j}(x^{\mu}-i\bar \theta \gamma^{\mu} \theta)\,,
\label{wtpi4}
\eeq
where\,\footnote{In the Euclidean space $\bar\psi$ becomes an independent variable.}
\beqn
&&
x^{\mu}=\{t,z\},\qquad \bar \theta=\theta^{\dagger}\gamma^{0},\qquad \bar \psi=\psi^{\dagger}\gamma^{0}
\nonumber
\\[2mm]
&&\gamma^{0}=\gamma^t=\sigma_2\,,\qquad \gamma^{1}=\gamma^z = i\sigma_1\,,\qquad \gamma_{5} 
\equiv\gamma^0\gamma^1 = \sigma_3\,.
\label{wtpi5}
\eeqn
With no twisted mass the Lagrangian  is \cite{Bruno}
(see also \cite{WessBagger})
\begin{equation}
\label{eq:kinetic}
{\cell}_{m=0}= \int d^{4 }\theta K(\Phi, \bar\Phi)
=G_{i\bar j} \left[\partial^\mu \bar\phi^{\,\bar j}\, \partial_\mu\phi^{i}
+i\bar \psi^{\bar j} \gamma^{\mu} \cde_{\mu}\psi^{i}\right]
-\frac{1}{2}\,R_{i\bar jk\bar l}\,(\bar\psi^{\bar j}\psi^{i})(\bar\psi^{\bar l}\psi^{k}).
\end{equation}
where
\beq
G_{i\bar j}=\frac{\partial^{2} K(\phi,\,\bar\phi)}{\partial \phi^{i}\partial \bar\phi^{\,\bar j}}
\label{wtpi6}
\eeq
 is the K\"ahler metric, and
$R_{i\bar jk\bar l}$ is the Riemann tensor \cite{Helgason}, 
\beq
R_{i\bar{j} k\bar{m}} = - \frac{g_0^2}{2}\left(G_{i\bar{j}}G_{k\bar{m}} +
G_{i\bar{m}}G_{k\bar{j}}
\right)\,.
\label{640}
\eeq
Moreover,
$$ \cde_{\mu}\psi^{i} ~~=~~
\partial_{\mu}\psi^{i} ~+~ \Gamma^{i}_{kl}\partial_{\mu} \phi^{k}\psi^{l}\,,
\qquad\qquad
	\Gamma^i_{kl} ~~=~~ -\, \frac{\delta^i_{\ (k} \delta_{l)\bi}\,\ov{\phi}{}^\bi}{\chi}
$$
is the covariant derivative.
The Ricci tensor $R_{i\bar j}$ is proportional to the metric \cite{Helgason},
\beq
\label{eq:RG}
R_{i\bar{j}} = \frac{g_{0}^2}{2}\,  N \, G_{i\bar{j}}\,.
\eeq
For the massless CP($N\!-\!1)$ model 
a particular choice of the K\"ahler potential
\begin{equation}
\label{eq:kahler}
K_{m=0}=\frac{2}{g_{0}^{2}}\ln\left(1+\sum_{i,\bar j=1}^{ N-1}\bar\Phi^{\,\bar j}\delta_{\bar j i}\Phi^{i}\right)
\end{equation}
corresponds to the round Fubini--Study metric.

Let us  briefly remind how one can introduce the twisted mass parameters \cite{twisted, Dor}.
The theory (\ref{eq:kinetic}) can be interpreted as an ${\mathcal N}=1$ theory of $N-1$ chiral superfields 
in four dimensions.  The theory possesses $N-1$ distinct   U(1) isometries 
parametrized by $t^{a}$, $a=1,\ldots,N-1$.
The Killing vectors of the isometries can be expressed via derivatives of the Killing 
potentials $D^{a}(\phi, \bar\phi)$,
\begin{equation}
\label{eq:KillD}
\frac{{d}\phi^{i}}{{  d}\,t_{a}}=-iG^{i\bar j}\,\frac{\partial D^{a}}{\partial \bar\phi^{ \,\bar j}}
\,,\qquad 
\frac{{d}\bar\phi^{ \,\bar j}}{{  d}\,t_{a}}=iG^{i\bar j}\,\frac{\partial D^{a}}{\partial \phi^{i}}\,.
\end{equation}
This defines the U(1) Killing potentials, up to additive constants.

The $N -1$ isometries are
 evident from the expression (\ref{eq:kahler}) for the K\"ahler potential, 
\begin{equation}
\label{eq:iso}
\delta\phi^{i}=-i\delta t_{a} (T^{a})^{i}_{k}(\phi)^{k}\,,\qquad 
\delta\bar\phi^{\,\bar j}=i\delta t_{a}(T^{a})^{\bar j}_{\bar l}\bar\phi^{\,\bar l}\,,
\qquad a=1,\ldots, N-1\,,
\end{equation}
(together with the similar variation of fermionic fields),
where the  generators $T^{a}$ have a simple diagonal form,
\begin{equation}
(T^{a})^{i}_{k}=\delta^{i}_{a}\delta^{a}_{k}\,, \qquad a=1,\ldots,N-1\,.
\end{equation}
 The explicit form of the Killing potentials $D^{a}$ in CP$(N\!-\!1)$ with the Fubini--Study metric is
\beq
\label{eq:KillF}
D^{a}=\frac{2}{g_{0}^{2}}\,\frac{\bar\phi\, T^{a}\phi}{1+\bar\phi\,\phi}\,,
\qquad a=1,\ldots,N-1\,.
\eeq
Here we use the matrix notation implying that $\phi$ is a column $\phi^{i}$ and 
$\bar\phi$ is a row $\bar\phi^{ \bar j}$.

The isometries allow us  to introduce an interaction with $N-1$ {\em external} 
U(1) gauge 
superfields $V_{a}$ by modifying, in a gauge invariant way,  the K\"ahler potential (\ref{eq:kahler}),
\begin{equation}
\label{eq:mkahler}
K_{m=0}(\Phi, \bar\Phi)\to
K_{m}(\Phi, \bar\Phi,V)\,.
\end{equation}
For CP$(N\!-\!1)$ this modification takes the form
\begin{equation}
\label{eq:mkahlerp}
K_{m}=\frac{2}{g_{0}^{2}}\ln \left(1+\bar\Phi\,{\rm e}^{V_{a}T^{a}}\Phi\right)\,.
\end{equation}
In every gauge multiplet $V_{a}$ let us retain only the $A^{a}_{x}$ and $A^{a}_{y}$ 
components of the gauge potentials taking them to be just constants,
\beq
V_{a}=-m_{a}\bar \theta(1+\gamma_{5})\theta -\bar m_{a}\bar \theta(1-\gamma_{5})\theta\,,
\label{wtpi1}
\end{equation}
where we introduced complex masses  $m_{a}$ as linear combinations of 
constant U(1) gauge potentials,
\beqn
m_{a}
&=&
A^{a}_{y}+iA^{a}_{x}\,,\qquad \bar m_{a}=m_{a}^{*}=A^{a}_{y}-iA^{a}_{x}\,,
\nonumber\\[2mm]
a 
&=&
 1,2, ..., N-1\,.
\label{wtpi2}
\eeqn
The introduction of the twisted masses does not 
break \ntwot supersymmetry in two dimensions.  To see this one can note that the mass parameters 
can be viewed as the lowest components of the twisted chiral superfields
$D_{2}\bar D_{1}V_{a}$.

Now we can go back to two dimensions implying that there is no dependence
on $x$ and $y$ in the chiral fields.  It gives us the Lagrangian with the twisted masses 
included \cite{twisted, Dor}:
\beqn
\label{eq:mtwist}
{\mathcal L}_{m}
&=&
 \int d^{4 }\theta \,K_{m}(\Phi, \bar\Phi,\,V)
=G_{i\bar j}\, g_{MN}\left[ {\cde}^M \bar\phi^{\,\bar j}\, {\cde}^{N} \phi^{i}
+i\,\bar \psi^{\bar j} \gamma^{M}\,D^{N} \psi^{i}\right]
\nonumber\\[3mm]
&-&
\frac{1}{2}\,R_{i\bar jk\bar l}\,(\bar\psi^{\bar j}\psi^{i})(\bar\psi^{\bar l}\psi^{k})\,,
\eeqn
where $G_{i\bar j} =\partial_{i}\partial_{\bar j}K_{m}|_{\theta=\bar\theta=0}$ is the K\"ahler metric 
and summation over $M$ includes, besides $M=\mu=0,1$, also 
$M=+,-$. 
The  metric $g_{MN}$ and extra gamma-matrices are
\begin{equation}
\label{eq:metric}
g_{MN}=\left(\begin{array}{crrr}1& 0& 0 & 0 \\0 & -1 & 0 & 0 \\[1mm]0 & 0 & 0 & -\frac 1 2 \\[1mm]0 & 0 & -\frac 1 2 & 0\end{array}\right),\qquad
\gamma^{+}=-i(1+\gamma_{5})\,,\quad
\gamma^{-}=i(1-\gamma_{5})\,.
\end{equation}
The gamma-matrices satisfy the following algebra:
\beq
\bar\Gamma^{M}\Gamma^{N}+\bar\Gamma^{N}\Gamma^{M}=2 g^{MN}\,,
\eeq
where the set $\bar\Gamma^{M}$ differs from $\Gamma^{M}$  by interchanging of
the $+,-$ components, $\bar\Gamma^{\pm}=\Gamma^{\mp}$.
The gauge covariant derivatives ${\cal D}^M$ are defined as
\beqn
&&
{\mathcal D}^{\mu}\phi=\partial^{\mu}\phi\,,\qquad {\cal D}^{+}\phi=-\bar m_{a}T^{a}\phi\,,
\qquad  {\cal D}^{-}\phi=m_{a}T^{a}\phi\,,
\nonumber\\[1mm]
&& {\cal D}^{\mu}\bar\phi
=\partial^{\mu}\bar\phi\,,
\quad ~{\cde}^{+}\bar\phi =\bar\phi\, T^{a}\bar m_{a}\,,
\qquad  {\cde}^{-}\bar\phi =-\bar \phi\, T^{a} m_{a}\,,
\nonumber\\
\eeqn
and similarly for ${\cal D}^{M}\psi$, while the general covariant derivatives $D^{M}\psi$ are
\begin{equation}
D^{M}\psi^{i}=
{\cal D}^{M}\psi^{i}+\Gamma^{i}_{kl}\,{\cal D}^{M}\! \phi^{k}\,\psi^{l}\,.
\end{equation}

\subsection{Geometric formulation, \boldmath{$\tilde\gamma \neq 0$}} 
\label{gftgnz}

The parameter of the heterotic deformation in the geometric formulation will be denoted by
$\tilde \gamma_\ssm$ (the tilde appears here for historical reasons;
perhaps, in the future it will be reasonable to omit it;
subscript $\scriptstyle (M)$ will stress that this section works with Minkowski notations).

To obtain the Lagrangian of the heterotically deformed model
we act as follows \cite{BSY3}: we start from (\ref{eq:kinetic}), add the
 right-handed spinor field $\zeta_R$, with the same kinetic term as in Sect.~\ref{gfsothd}, and
 add the bifermion terms
 \beq
 \frac{g_0}{\sqrt 2}
 \left[ \tilde{\gamma}_\ssm\, \zeta_RG_{i\bar j}\left(i\pt_L\bar\phi^{\,\bar j}\right)\psi_R^i + 
        \bar{\tilde \gamma}{}_\ssm\, \bar\zeta_R G_{i\bar j}\left(i\pt_L\phi^{i}
 \right)\bar\psi_R^{\,\bar j} 
 \right].
 \eeq
Next, we change the four-fermion terms exactly in the same way this was done in
 \cite{SY1},
 namely
 \beqn
&-&\frac{1}{2} \,R_{i\bar jk\bar l}\left[\left(\bar\psi^{\bar j}\psi^{i}\right)\left(\bar\psi^{\bar l}\psi^{k}\right)
\left(\bar\psi^{\bar j}\psi^{i}\right)\left(\bar\psi^{\bar l}\psi^{k}\right)
\right]
~\longrightarrow~
\nonumber\\[3mm]
&-& \frac{g_0^2}{2}\left( G_{i\bar j}\psi^{\dagger\, \bar j}_R\, \psi^{ i}_R\right)
\left( G_{k\bar m}\psi^{\dagger\, \bar m}_L\, \psi^{ k}_L\right)+
\frac{g_0^2}{2}\left(1-|\tilde\gamma_\ssm|^2\right)
\left( G_{i\bar j}\psi^{\dagger\, \bar j}_R\, \psi^{ i}_L\right)
\left( G_{k\bar m}\psi^{\dagger\, \bar m}_L\, \psi^{ k}_R\right)\,
\nonumber
\\[4mm]
&-&
 \frac{g_0^2}{2} \, |\tilde{\gamma}_\ssm |^2 \,\left(\zeta_R^\dagger\, \zeta_R
\right)\left(G_{i\bar j}\,  \psi_L^{\dagger\,\bar j}\psi_L^i\right),
\eeqn
 where the first line represents the last term in Eq.~(\ref{eq:kinetic}), and we used the identity~(\ref{640}).
 If one of the twisted masses from the set  $\{m_1,\,m_2,\, ..., m_N\}$ vanishes (say, $m^N=0$),
then this is the end of the story. The masses 
$m_a$ in Eqs. (\ref{wtpi1}) and (\ref{wtpi2}) are $\{m_1,\,m_2,\, ..., m_{N-1}\}$.

However, with more general twisted mass sets,
for instance, for the \zn-symmetric masses (\ref{two}), 
one arrives at a more contrived situation since 
one should take into account an extra contribution.
Occurrence of this contribution can be seen \cite{BSY3} in a relatively concise
manner using the superfield formalism of \cite{SY1},
\beq
\Delta\cell \sim M\int  {\mathcal B}\,\,  d\bar\theta_L \, d\theta_R + {\rm H. c.}\,,
\label{tftpi1}
\eeq
where
${\mathcal B}$ is a (dimensionless) \nzt superfield\,\footnote{This means that ${\mathcal B}$
is the superfield only with respect to the right-handed transformations.}
\beq
{\mathcal B} =\left\{\zeta_R\left(x^\mu + i\bar\theta\gamma^\mu\theta\right) +\sqrt{2} \theta_R {\mathcal F}
\right\} \bar\theta_L\,.
\label{tftpi2}
\eeq
 The parameter $M$ appearing in (\ref{tftpi1}) has dimension of mass; in fact, it is proportional to $m^N$.
 
 As a result, the heterotically deformed \cpn Lagrangian with all $N$ twisted mass parameters included
can be written in the following general form:
 \beq
 \cell = \cell_\zeta + \cell_{m=0} +\cell_m\,,
 \label{tftpi3}
 \eeq
 where the notation is self-explanatory. The expression for $\cell_m$ is quite cumbersome.
 We will not reproduce it here, referring the interested reader to \cite{BSY3}.
 For convenience, we present here the first two terms,
\beqn
\cell_\zeta + \cell_{m=0} && 
= 
\zeta_R^\dagger \, i\partial_L \, \zeta_R  + 
\left[\tilde\gamma_\ssm\, \frac{g_0}{\sqrt 2} \, \zeta_R  \, 
      G_{i\bar j}\,  \big( i\,\partial_{L}\phi^{\dagger\,\bar j} \big)\psi_R^i
      + {\rm H.c.}\right]
\nonumber
\\[4mm]
&&
 -\frac{g_0^2}{2}\, |\tilde\gamma_\ssm |^2 \,\left(\zeta_R^\dagger\, \zeta_R
\right)\left(G_{i\bar j}\,  \psi_L^{\dagger\,\bar j}\psi_L^i\right)
\nonumber
\\[4mm]
&&
+G_{i\bar j} \big[\partial_\mu \phi^{\dagger\,\bar j}\, \partial_\mu\phi^{i}
+i\bar \psi^{\bar j} \gamma^{\mu} D_{\mu}\psi^{i}\big]
\nonumber
\\[4mm]
&&
- \frac{g_0^2}{2}\left( G_{i\bar j}\psi^{\dagger\, \bar j}_R\, \psi^{ i}_R\right)
\left( G_{k\bar m}\psi^{\dagger\, \bar m}_L\, \psi^{ k}_L\right)
\nonumber
\\[4mm]
&&
+\frac{g_0^2}{2}\left(1-|\tilde\gamma_\ssm|^2\right)
\left( G_{i\bar j}\psi^{\dagger\, \bar j}_R\, \psi^{ i}_L\right)
\left( G_{k\bar m}\psi^{\dagger\, \bar m}_L\, \psi^{ k}_R\right)\,,
\label{cpn-1g}
\eeqn
where we used (\ref{640}). The above Lagrangian is \nzt-supersymmetric at the classical level.
Supersymmetry is spontaneously broken by nonperturbative effects \cite{EdTo,SYhet}. 
Inclusion of $\cell_m$ spontanesously breaks supersymmetry at the classical level (see
Eq.~(\ref{ferbosmasssplit}) and 
Eq.~(2.11) in \cite{BSY3}).

The relation between $\tilde\gamma$ and $\delta$
is as follows \cite{BSY3}:
\beq
  i\,\tilde\gamma_\ssm =
     \tilde\gamma_{\scriptscriptstyle(E)} = \sqrt{2} \frac{\delta}{\sqrt{1+ 2 |\delta |^2}},
\label{tftpi4}
\eeq
implying that $\tilde \gamma$ does {\em not} scale with $N$ in the 't Hooft limit.
See Appendix~\ref{app:mink} for details on relation between Euclidean and Minkowski notations.

\end{document}